\documentclass[a4paper]{jpconf}
\usepackage{graphicx}

\begin{document}

\newcommand{\Msun}{M_\odot}
\newcommand{\be}{\begin{equation}}
\newcommand{\ee}{\end{equation}}
\newcommand{\bea}{\begin{eqnarray}}
\newcommand{\eea}{\end{eqnarray}}
\newcommand{\LCDM}{$\Lambda$CDM }

\vspace*{-3cm}\hspace{11.3cm}IFT--UAM/CSIC--17--016

\title{Massive Primordial Black Holes as Dark Matter and their detection with Gravitational Waves}

\author{Juan Garc\'ia-Bellido}

\address{Instituto de F\'isica Te\'orica UAM-CSIC, Universidad Aut\'onoma de Madrid, Cantoblanco, 28049 Madrid, Spain}

\ead{juan.garciabellido@uam.es}

\begin{abstract}
Massive Primordial Black Holes (MPBH) can be formed after inflation due to broad peaks in the primordial curvature power spectrum that collapse gravitationally during the radiation era, to form clusters of black holes that merge and increase in mass after recombination, generating today a broad mass-spectrum of black holes with masses ranging from 0.01 to $10^5~\Msun$. These MPBH could act as seeds for galaxies and quick-start structure formation, initiating reionization, forming galaxies at redshift $z>10$ and clusters at $z>1$. They may also be the seeds on which SMBH and IMBH form, by accreting gas onto them and forming the centers of galaxies and quasars at high redshift. They form at rest with zero spin and have negligible cross-section with ordinary matter. If there are enough of these MPBH, they could constitute the bulk of the Dark Matter today. Such PBH could be responsible for the observed fluctuations in the CIB and X-ray backgrounds. MPBH could be directly detected by the gravitational waves emitted when they merge to form more massive black holes, as recently reported by LIGO. Their continuous merging since recombination could have generated a stochastic background of gravitational waves that could eventually be detected by LISA and PTA. MPBH may actually be responsible for the unidentified point sources seen by Fermi, Magic and Chandra. Furthermore, the ejection of stars from shallow potential wells like those of Dwarf Spheroidals (DSph), via the gravitational slingshot effect, could be due to MPBH, thus alleviating the substructure and too-big-to-fail problems of standard collisionless CDM. Their mass distribution peaks at a few tens of $\Msun$ today, and could therefore be detected also with long-duration microlensing events, as well as by the anomalous motion of stars in the field of GAIA. Their presence as CDM in the Universe could be seen in the time-dilation of strong-lensing images of quasars. The hierarchical large scale structure behaviour of MPBH does not differ from that of ordinary CDM.
\end{abstract}

\section{Introduction}\label{sec:Intro}

\

The discovery of several massive black hole mergers by the LIGO interferometer~\cite{Abbott:2016blz,Abbott:2016nmj,Abbott:2016drs,TheLIGOScientific:2016wyq,TheLIGOScientific:2016htt,TheLIGOScientific:2016pea} has opened the new era of Gravitational Wave Astronomy, exactly one century after Albert Einstein proposed his theory of gravitation. The masses of these black holes are somewhat larger than expected from remnants of supernovae explosions and stellar evolution~\cite{TheLIGOScientific:2016htt,Belczynski:2016obo}, so it is possible that LIGO has discovered a whole new population of massive black holes, formed in the early universe. Our aim is to understand how to generate such a population, and search for any signatures that will distinguish them from stellar black holes. It was first proposed in Ref.~\cite{GarciaBellido:1996qt} that massive primordial black holes (MPBH) could arise from high peaks in the primordial matter power spectrum that collapse gravitationally during the radiation era. More recently, we found~\cite{Clesse:2015wea} that broad peaks in the power spectrum would form clusters of black holes, which would merge and accrete mass (gas and smaller black holes) since recombination, reaching tens to hundreds (a few of them can reach billions) of solar masses today. These MPBH could be the dominant of the Dark Matter in the Universe~\cite{GarciaBellido:1996qt}; they would act as seeds for the first galaxies, reionizing the intergalactic medium and forming galaxies and quasars at high redshift~\cite{Clesse:2015wea}. Here we will explore the consequences of this new scenario for the understanding of dark matter and extend the very rich phenomenology that primordial black holes induce in the late Universe.

Since Zwicky proposed it to explain the stability of clusters of galaxies~\cite{Zwicky:1933gu}, the nature of dark matter has puzzled both physicists and astronomers alike. Today we have evidence of dark matter from the rotation curves of galaxies~\cite{Rubin:1970zza}, from gravitational lensing on large scale structures~\cite{Massey:2007wb} and from the temperature anisotropies in the CMB~\cite{Ade:2015xua}. According to the usual $\Lambda$CDM paradigm, structures like galaxies and clusters of galaxies form by hierarchical accretion of ordinary gas on dark matter halos. All observations so far rely solely on the gravitational interaction of dark matter, but gives no hint about its other interactions. Many different hypothesis have been put forward on its nature, with ninety orders of magnitude in the range of masses of its possible components, from ultralight axions~\cite{Hui:2016ltb} to massive black holes~\cite{Bertone:2010zza}. Until recently, the most popular candidates have been some weakly interacting particles (WIMP) predicted in certain extensions of the standard model of particle physics, but no evidence of any such particle has been found so far at the Large Hadron Collider (LHC) at CERN, neither in underground facilities, with cryogenic detectors like CDMS or liquid Xenon detectors like LUX, nor by indirect searches of their annihilation with astronomical satellites like Fermi~\cite{Bertone:2010zza}.

An alternative and very natural hypothesis, considered for a long time, is that of MAssive Compact galactic Halo Objects (MACHOs), such as planets, brown dwarves, neutron stars or black holes. Such compact objects were searched for decades ago, thanks to the microlensing of millions of stars~\cite{Paczynski:1985jf,Paczynski:1991zz} in nearby galaxies like the Large Magellanic Cloud (LMC) or Andromeda~\cite{MACHO,EROS,OGLE,LMC,Allsman:2000kg,AGAPE,Tisserand:2006zx,Wyrzykowski:2015ppa}. The search stopped ten years ago when the experiments failed to find MACHOs below 10 solar masses as the main component of the dark matter halo of our galaxy~\cite{LMC}, since compact objects with larger masses were unexpected~\cite{Allsman:2000kg}. 
Stellar black holes are the end points of stellar evolution under gravitational collapse after a supernova explosion. Most of these black holes have stellar masses, but only a few are above $10~\Msun$. Their masses are usually determined from the X-ray emission of their accretion disks. The more massive BH are supposed to arise from the gravitational collapse of population III supermassive stars, with very low metallicity, otherwise most of their mass will be blown away and will not end up inside the black hole. Nevertheless, it is difficult to produce BH from stellar evolution with masses above $10~\Msun$, even in the progenitor star had a mass above $250~\Msun$~\cite{Belczynski:2016obo}. It is therefore rather surprising that the first direct detection of gravitational waves from the merger of two black holes involved not just one but two $\sim 30~\Msun$ black holes. So, how did they form in the first place? and second, how did they find each other and merge within the lifetime of the Universe? 

\begin{figure}[ht]
    \begin{center}
        \includegraphics[width=10cm]{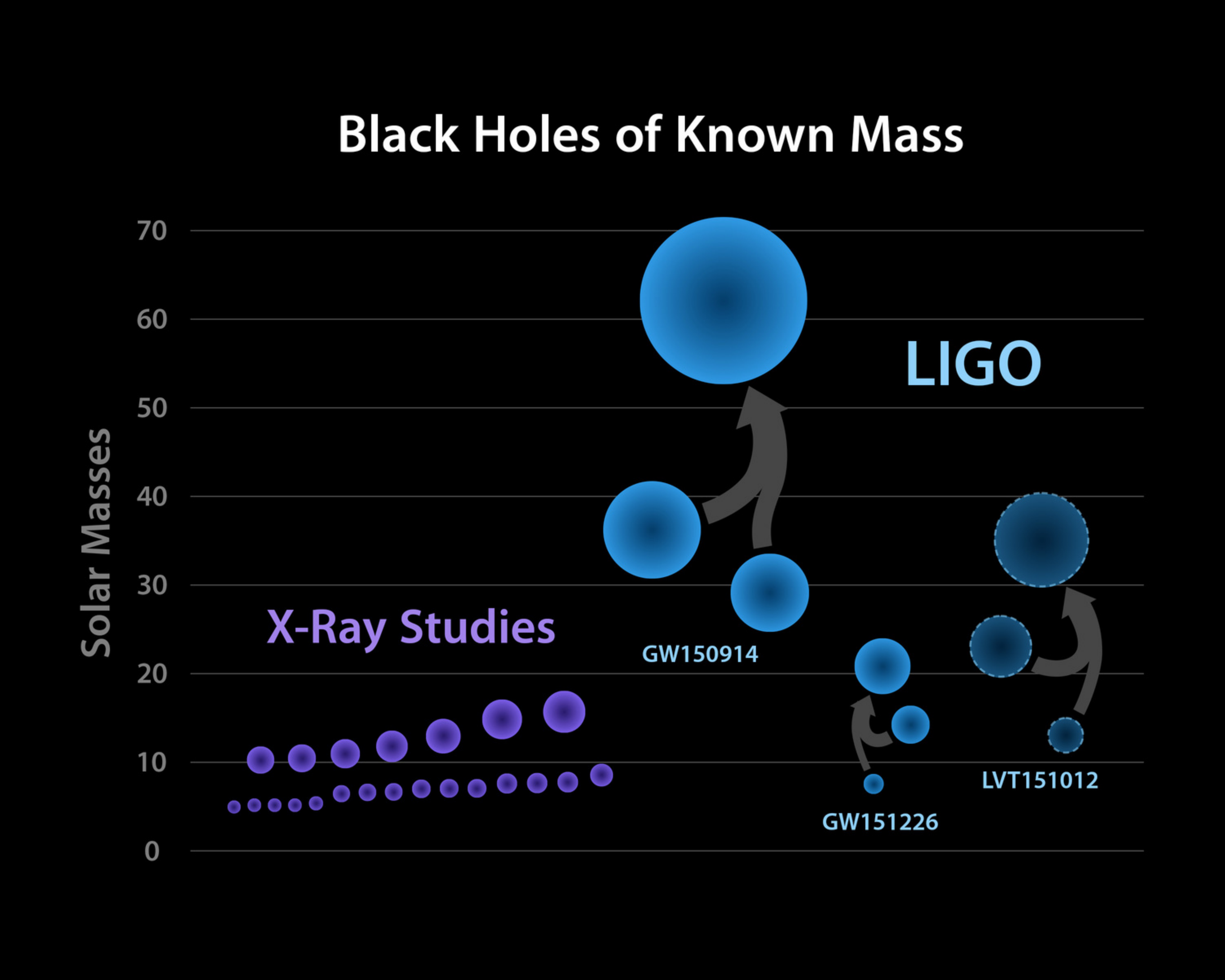}
    \end{center}
    \caption{The Black Holes of Known Mass detected by LIGO. It is clear that they correspond to a new population of black holes unheard off before. While IMBH and SMBH were known to populate the centers of globular clusters and galaxies, respectively, this new class of black holes in binaries had not been detected before. Figure from [LIGO webpage].} 
    \label{fig:BHM}
\end{figure}

With the recently born gravitational wave astronomy thanks to LIGO and the first detection of $30~\Msun$ black holes, we have opened again the window on the search for MPBH as the main component of dark matter in the Universe~\cite{Clesse:2015wea,Bird:2016dcv,Clesse:2016vqa,Sasaki:2016jop,Carr:2016drx}. Furthermore, massive PBH could be the missing link between early star and galaxy formation soon after recombination and the supermassive black holes at the centers of quasars and active galactic nuclei observed at high and intermediate redshifts. The use of gravitational wave interferometers as astronomical tools will help characterize this new population of black holes and measure their mass and spin distribution, which will allow astronomers to identify their progenitors and formation rates.

\begin{figure}[ht]
    \begin{center}
    \hspace*{-13mm}
        \includegraphics[width=13.6cm]{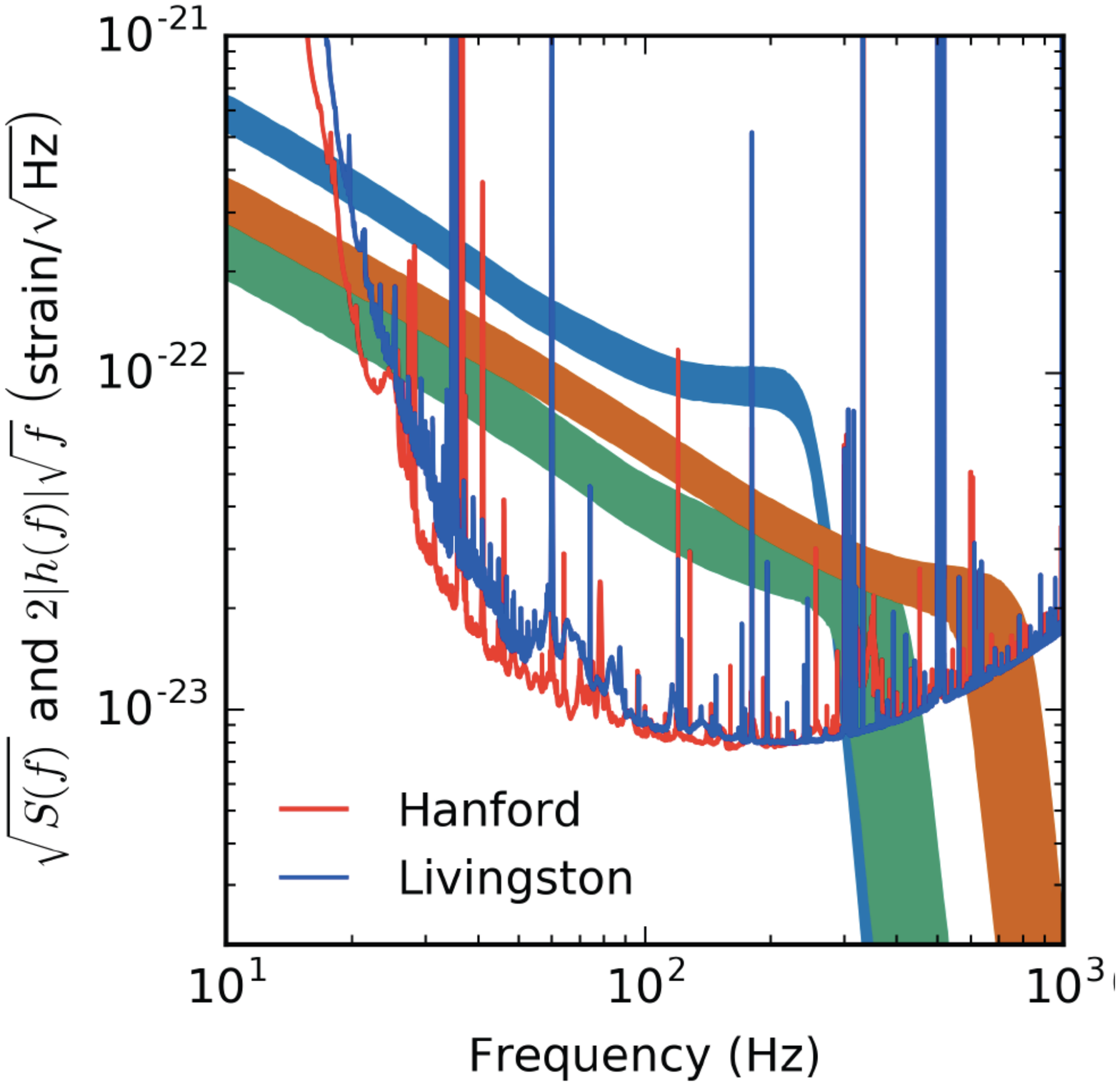} 
    \hspace*{11mm}
        \includegraphics[width=13.4cm]{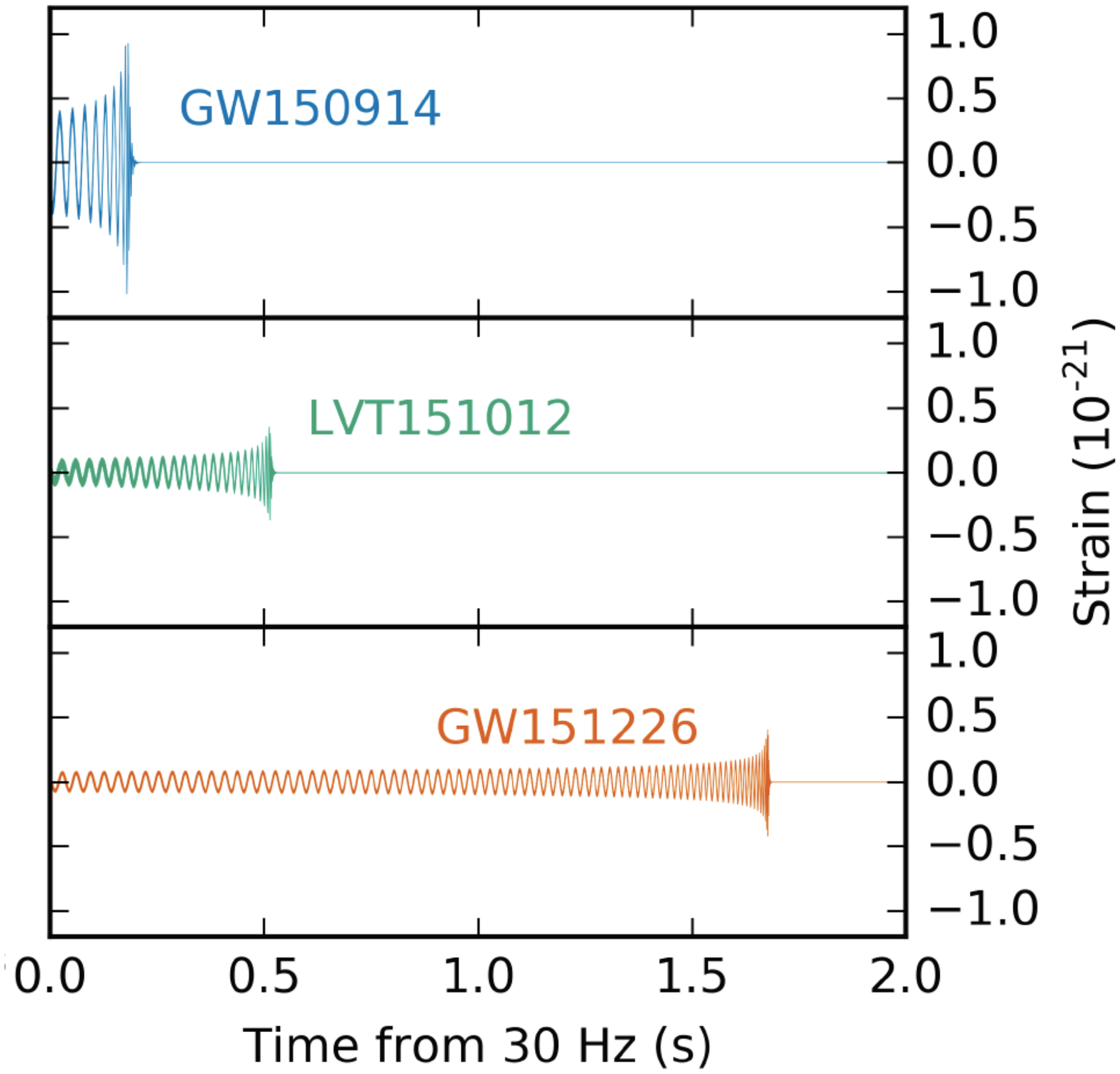}
    \end{center}
    \vspace*{-7mm}
    \caption{The strain sensitivity and amplitude of the three GW events detected by advanced LIGO as they inspiral towards merging. The inspiralling binary black hole (BBH) coalescence time-span after the 30 Hz mark of the three events. More massive binaries would not be seen by LIGO, since $f_{\rm ISCO} = 44\ {\rm Hz}~(100~\Msun/M_{\rm tot}) > 30$~Hz implies $M_{\rm tot} < 147~\Msun$, but better seismic attenuation by Virgo and KAGRA may help explore the higher mass range. Figures from Ref.~\cite{TheLIGOScientific:2016pea}.} 
    \label{fig:LIGOevents}
\end{figure}

In fact, LIGO has detected not just one massive BH binary merger, but {\em three} events, in the course of run O1, from September 2015 to January 2016, see Figs.~\ref{fig:BHM} and~\ref{fig:LIGOevents}. The masses of the inspiralling BH before merging are not all equal, they range from 8 to $36~\Msun$, and the final BH masses range from 23 to $62~\Msun$. If there is a population of BHs out there with large masses, which do not arise from stellar evolution, where did they come from?

A natural hypothesis is that these massive black holes are of primordial origin. That is, they were there before the stars started to shine. The first proposal of primordial black holes dates back to Hawking's work on black hole evaporation. He expected tiny black holes, of masses of order $10^{15}$~g, to be evaporating today, emitting in their last stages a burst of energy that he thought at the time could be the observed gamma ray bursts~\cite{Carr:1974nx,Carr:1975qj}. Now we know that GRB have a different origin and, moreover, these PBH are too light to account for the LIGO observations. Since then, a series of proposals have been made. For instance, if the QCD quark-hadron transition were first order, then the resulting zero-pressure fluid may have allowed primordial fluctuations to collapse to form black holes below~\cite{Dolgov:1992pu} and around a solar mass, corresponding to the mass within the horizon at that time. This was further studied by Jedamzik~\cite{Jedamzik:1996mr,Jedamzik:1999am}, to account for the measured microlensing events towards the LMC, found by the MACHO collaboration~\cite{LMC}. Now we know that the QCD transition is nor first order, but actually is a crossover, so the equation of state is non-zero, and the small fluctuations generated during inflation cannot give rise to PBH at that stage.

Our scenario of massive PBH is a different scenario, based on the fundamental physics of inflation, which has important consequences for the whole of Astronomy and Cosmology. We would like to understand the nature of these massive black holes detected by LIGO and determine their origin, whether primordial or not. If the merging rate is as high as the LIGO collaboration seem to suggest, then we will have hundreds of events in the next few years, enough to convince ourselves that we are in front of a new class of BH. This new scenario opens up a multi-probe, multi-epoch and multi-wavelength approach to the nature of Dark Matter in the form of MPBH. 

In this brief review we explore the very rich phenomenology that arises in this scenario, and describe a broad range of new signatures that could be measured in a wide variety of astrophysical and cosmological observations in order to test this hypothesis. In section {\bf II}, I describe the various formation scenarios, with different ways of generating peaks in the matter power spectrum, studying the different models of PBH growth via gas accretion and BH merging. In section {\bf III}, I describe the very rich phenomenology that such a scenario puts forward and connects many different aspects of Astrophysics and Cosmology. It is possible that in the next decade we may have a final picture of the nature of dark matter. In section {\bf IV}, I explore one particular aspect of the whole list of phenomenological signatures, that of the anomalous motion of stars in the solar neighbourhood, which may be detectable in the near future by the GAIA mission, and entertain on the phenomena of gravitational slingshot for ejecting stars from Globular Clusters and Dwarf Spheroidals (DSph). In section {\bf V}, I describe evidence for microlensing in the time residuals of strong lensing of distant quasars. In section {\bf VI}, I explore the gamma and X-ray emissions from MPBH in our galaxy and local group as seen by Fermi and Chandra respectively. In section {\bf VII}, I study the different sources of GW from MPBH. In section {\bf VIII}, I conclude with a summary and some discussion.

\begin{figure}[ht]
    \begin{center}
        \includegraphics[width=12cm]{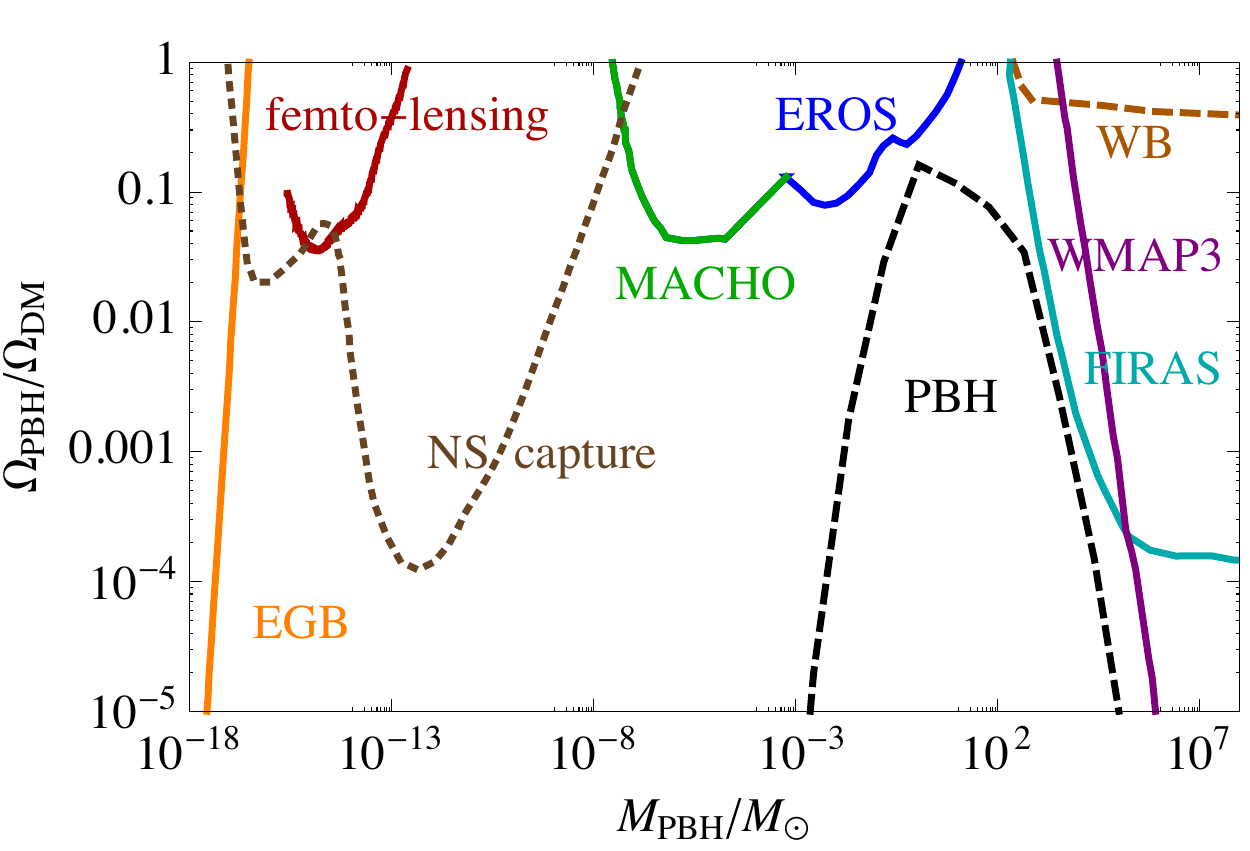}
    \end{center}
    \caption{Limits on the abundance of PBH today, from extragalactic photon background (orange), femto-lensing (red), micro-lensing by MACHO (green) and EROS (blue), from wide binaries (light brown), and CMB distortions by FIRAS (cyan) and WMAP3 (purple).  The constraints from star formation and capture by neutron stars in globular clusters are displayed for $\rho_{\rm DM}^{\rm Glob. Cl.} = 2 \times 10^3 \ {\rm GeV/cm}^{3}$ (dark brown). The  black dashed line corresponds to a particular realization of our scenario of PBH formation. Figure adapted from Ref.~\cite{Clesse:2015wea}.} 
    \label{fig:Constraints}
\end{figure}

\begin{figure}[ht]
    \begin{center}
        \includegraphics[width=12cm]{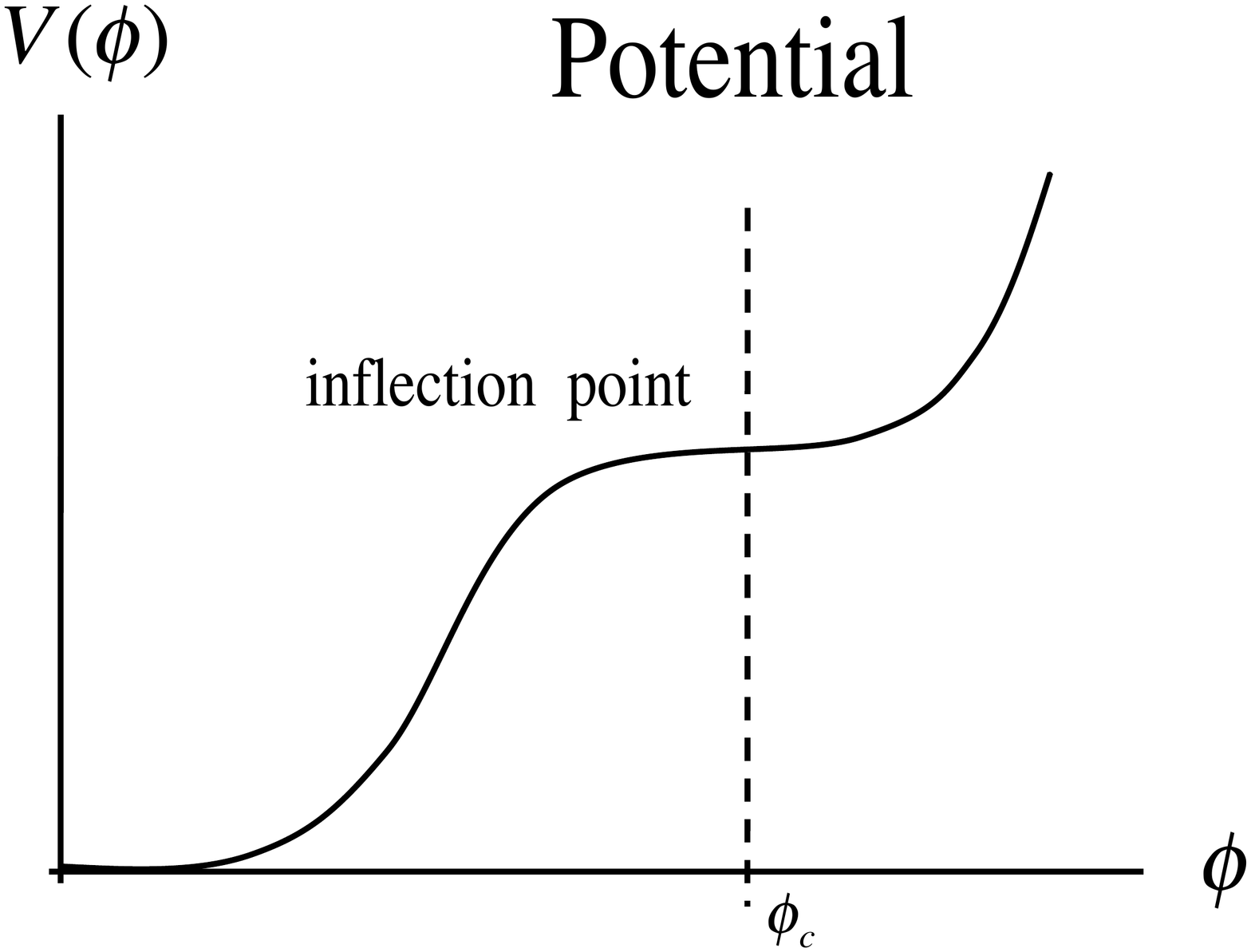}
    \end{center}
    \caption{Generic inflationary potential giving rise to peaks in the power spectrum. In two-field models like hybrid inflation the inflection point is substituted by the symmetry breaking point. In both cases there are a few-folds of inflation after the critical point $\phi_c$.} 
    \label{fig:Potential}
\end{figure}

\section{Massive PBH formation scenarios}
\label{sec:Form}

\

In 1996, together with Linde and Wands~\cite{GarciaBellido:1996qt}, we realized that, if the primordial curvature fluctuation spectrum had a pronounced peak at some particular scale, it was possible to generate PBHs whenever those high curvature fluctuations reentered the horizon during the radiation era, since nothing could prevent their gravitational collapse. The mass of those PBH would be given very approximately by the total mass within the horizon at the time of reentry. Moreover, in certain realizations of inflation, it was possible that the number of PBH would be so large that these could become the bulk of the dark matter in the universe~\cite{GarciaBellido:1996qt}. At the time, the constraints on BHs of a solar mass were not very strong. Thanks to MACHO and EROS collaborations, the window on solar mass BHs as the sole (100\%) contributor to the Milky Way halo was ruled out, and very stringent bounds were put at the level of less than 10\% of the halo in masses above 10~$\Msun$. This seemed to close the window on massive PBH as the origin of dark matter, and thus encouraged the search for particle dark matter (PDM) as the main contributor to the DM of the Universe. Some groups still consider the possibility of low-mass PBH (below planetary size) as the main component of DM~\cite{Frampton:2010sw}, although there are very stringent constraints on these, see e.g. the reviews Ref.~\cite{Khlopov:2008qy,Carr:2016drx}.

More recently, Clesse and Garc\'ia-Bellido proposed a scenario~\cite{Clesse:2015wea} where, rather than a single sharp peak in the matter power spectrum, one could generate a {\em broad spectrum} of PBH masses from a large peak in the primordial spectrum of curvature fluctuations from inflation. This broad mass spectrum could account for all of the dark matter, and nevertheless pass all the bounds from wide binaries and microlensing events of our halo, since any given range of masses is always below the 10\% fraction of the halo mass, see Fig.~\ref{fig:Constraints}.

%\be
%\ee

\begin{figure}[ht]
    \begin{center}
        \includegraphics[width=12cm]{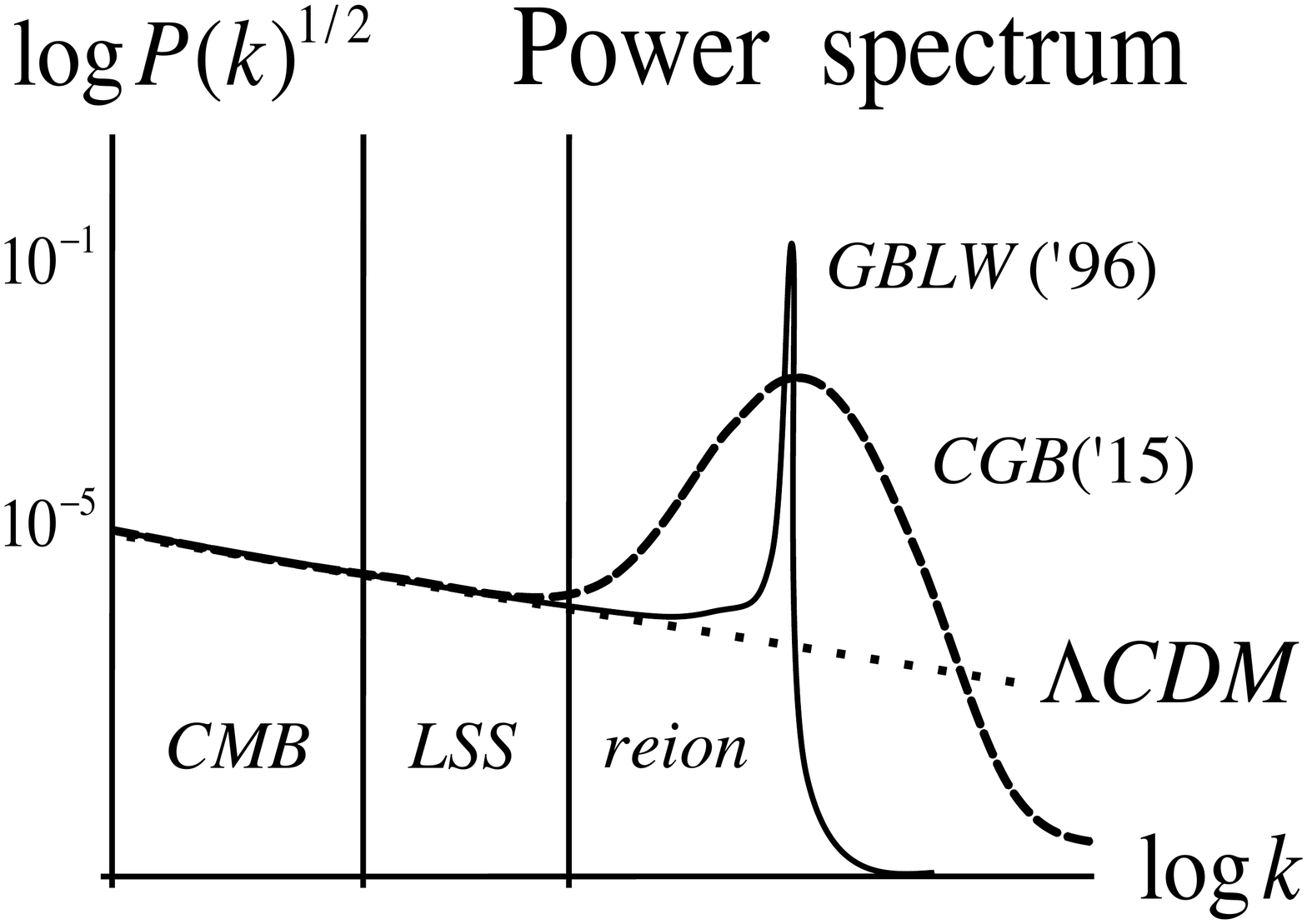} \\
        \includegraphics[width=12cm]{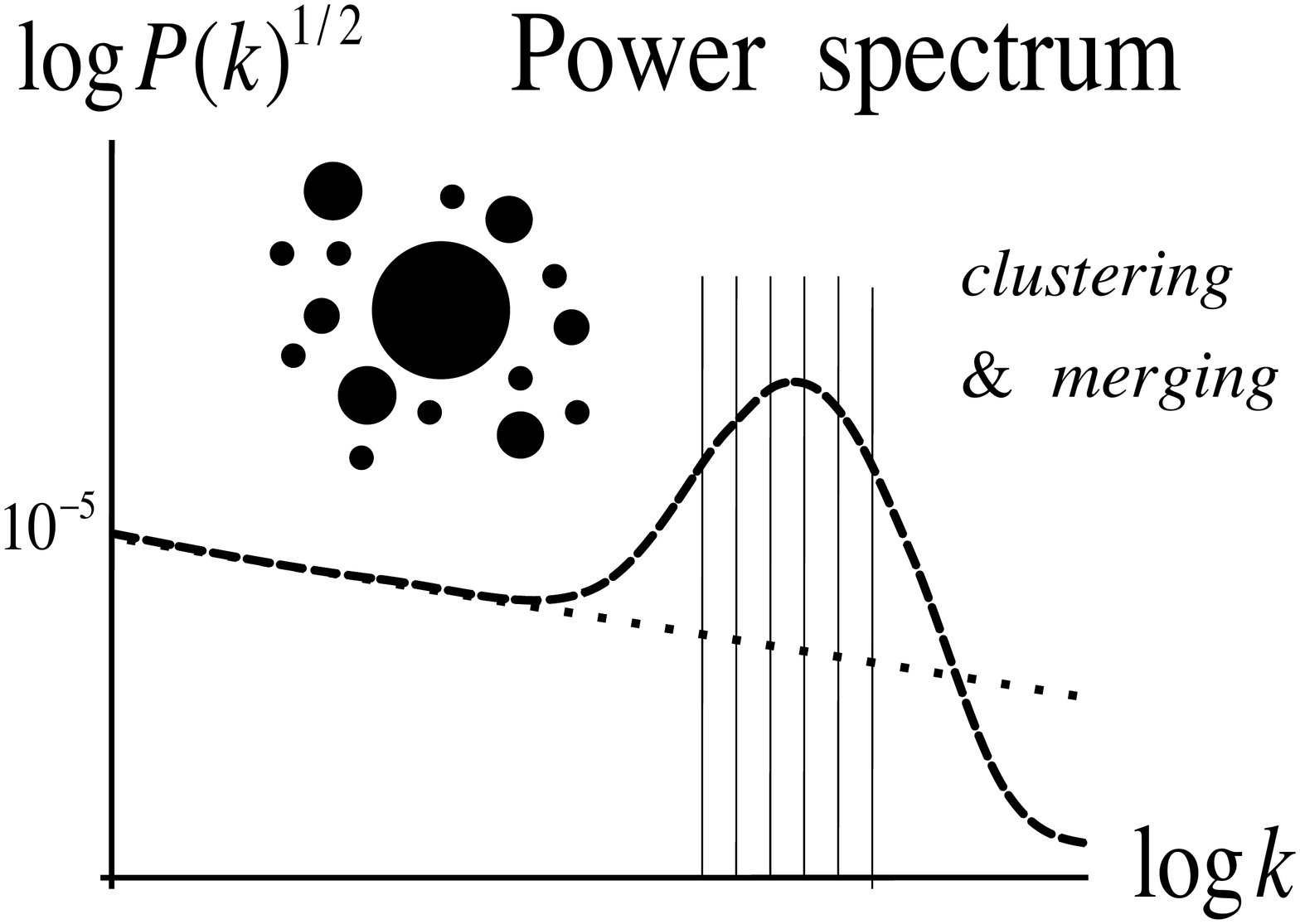}
    \end{center}
    \caption{The primordial power spectrum of curvature fluctuations induced during inflation. Those fluctuations that enter during the radiation era collapse to form black holes within a range of masses, and spatially clustered around the more massive ones. The dotted line corresponds to the Gaussian primordial spectrum predicted by inflation and at the core of the \LCDM paradigm, and consistent with the observed CMB anisotropies; the continuous line corresponds to the original model of GBLW (1996)~\cite{GarciaBellido:1996qt}, with a sharp peak in the spectrum, that gives rise to a monochromatic mass spectrum, and the dashed line corresponds to the recent proposal of CGB (2015)~\cite{Clesse:2015wea} for a broad peak in the spectrum, giving rise to a broad mass spectrum of PBH, which are strongly clustered.} 
    \label{fig:PowerSpectrum}
\end{figure}

\subsection{Peaks in the primordial Power Spectrum}

\

The width of the peak in the matter power spectrum that gives rise to the mass range of PBH depends on the particular inflationary model behind the origin of curvature fluctuations that gave rise to the formation of structure in the universe. Some models of inflation may have a phase transition 20 e-folds before the end of inflation, triggered by a symmetry breaking field, like in hybrid inflation~\cite{GarciaBellido:1996qt}, and this will give rise to a very pronounced peak in the spectrum, whose width will depend on the parameters of the model. Another possibility is to have the inflaton coupled to gauge fields, which grow exponentially and themselves backreact on the curvature, producing high peaks in the spectrum~\cite{Garcia-Bellido:2016dkw}. Alternatively, one can think of single-field models with an inflection point in the potential, see Fig.~\ref{fig:Potential}, which slows down the scalar field driving inflation and produces a broad peak in the spectrum~\cite{Garcia-Bellido:2017mdw}.

While the amplitude of anisotropies in the cosmic microwave background cannot exceed a few parts in $10^{-5}$ on the scale of the horizon, the fluctuation spectrum on much smaller scales is largely unconstrained. The CMB only probes the 50-to-60 efolds range, depending on the actual temperature of reheating after inflation, and future 21cm surveys with intensity mapping will probe the reionization epoch of 50-to-40 efolds range, see Fig.~\ref{fig:PowerSpectrum}. The peak in the primordial spectrum would begin to become important around 40-to-20 efolds before the end of inflation, on scales never directly probed by any astronomical survey, since on those scales, density fluctuation quickly become non-linear and gravitational collapse washes out any trace of the primordial spectrum. Nevertheless, if there is a large peak, as in Fig.~\ref{fig:PowerSpectrum}, the gravitational collapse of those fluctuations upon reentry will form a broad mass distribution of PBH, that could act as seeds for massive stars and galaxies.

Observations will soon determine the mass and spin spectrum of the MPBH responsible for the measured Gravitational Waves from BBH coalescence. This could then be used as a test of the mechanisms for PBH production, opening a new window into the last 40 e-folds of inflation, and thus help us access the ultra high energy fundamental physics domain that is completely inaccessible by particle physics accelerators~\cite{Garcia-Bellido:2017mdw}.

\subsection{Accretion and merging}
\label{sec:Merging}

\

In this subsection we study the evolution through accretion and merging of the population of PBH. On the one hand, during the radiation era, they can accrete gas and radiation via Bondi-Hoyle accretion but this rate is negligible. Later, during the matter era, they can start to accrete faster, but always below the Eddington limit. Alternatively, if they acquire an accretion disk of gas, and the outflows are along the axis, like in quasars and microquasars~\cite{Mirabel:1998hf}, they can accrete much faster. On the other hand, since MPBH originate in clusters, they can find each other and merge more easily, by gravitational attraction, emitting gravitational waves in the process, and generating a stochastic background~\cite{Clesse:2016ajp}.

Therefore, their masses grow at a significant rate, both through accretion and merging, thus shifting their mass distribution towards larger values since recombination, and increasing the high-mass tails of the distribution. On top of that, since PBH form a fluid of zero pressure, their density contrast (i.e. inhomogeneities in the spatial distribution of PBH, arising from the standard metric fluctuations induced during inflation) grows like any matter component, linear with the scale factor during the matter era. Apart from changes in their mass distribution, we expect also changes in their spin distribution. PBH are born without spin since they arise from the gravitational collapse of an isotropic gas upon horizon crossing of order-one fluctuations in curvature. However, the spin will build up from subsequent mergers. As PBH masses start to build up through merging of smaller PBH, they can acquire a significant spin from the angular momentum of the inspiralling orbit. However, subsequent mergers occur from a random distribution of orientations of the orbits, and it is likely that the final spin distribution will be centered around the zero-spin configuration, with a dispersion that depends on the merging history and mass distribution. We are performing numerical relativity simulations in order to derive a prediction for the mass and spin distributions of present black holes.

\section{Signatures of massive PBH}
\label{sec:Signatures}

\

In this section we will discuss the possible signatures of primordial black holes as the dominant component of Dark Matter. I will itemize here some of those signatures that I believe could be used to distinguish PBH from stellar black holes. They range from early to late epochs of the evolution of the Universe, with different probes (electromagnetic and gravitational waves) and different frequencies (from nHz to GHz and from meV to tens of GeV). Some of these signatures are ready to be explored; others will take a few years, or even decades. There is a lot of work ahead to thoroughly test the scenario.

\begin{itemize}

\item {\bf CMB distortions and anisotropies}. Massive PBH formation during the radiation era has important consequences on the CMB. Gas falling into PBH through Bondi-Hoyle accretion will reinject some fraction of its energy into the strongly-coupled plasma before recombination, which will then distort the blackbody spectrum of photons at decoupling~\cite{Mack:2006gz}. There are stringent limits on the CMB $y$ and $z$-distortions, which can be evaded if the PBH have small masses before recombination, which then grow during the matter era. The most stringent bounds come from a paper by Ricotti et al.~\cite{Ricotti:2007au}, which has an error in Eq.(44): There is a factor $(1+z)^2$ missing, which changes the bounds by several orders of magnitude towards larger PBH masses, leaving a range of values around $M\sim100\,\Msun$ unconstrained, see also~\cite{Ali-Haimoud:2016mbv,Blum:2016cjs}. This is the reason why we shifted these bounds (labeled FIRAS and WMAP) to the right in Fig.~\ref{fig:Constraints}.

\item {\bf First stars in the Universe}. The random distribution of PBH will act as seeds for gas to fall and initiate star formation and the reionization of the universe. This will generate a UV and gamma background at high redshift ($z\sim20$) that could be seen today redshifted into the infrared and soft X-ray, respectively. The recent measurement of strong cross-correlations between fluctuations in the CIB and the diffuse X-ray background~\cite{Kashlinsky:2016sdv} suggests that a population of PBH could have initiated star formation and reionization at high redshift and also be responsible for the sources generating the present X-ray background.

\item {\bf Reionization and 21cm Intensity mapping}. While present CMB experiments give us a measurement of the integrated optical depth from the last scattering surface to the present~\cite{Ade:2015xua}, future 21cm intensity mapping surveys like SKA will be able to measure the whole history of reionization, from redshift $z\sim20$ to $z\sim6$, thus constraining the local sources of reionization at high redshift~\cite{Blake:2004pb}. The power spectrum of fluctuations on 21cm maps will put very stringent limits on the mass distribution of PBHs.

\item {\bf Early galaxy formation}. A population of massive primordial black holes would act as centers of accretion of ordinary gas, and give rise to galaxies at high redshift ($z\sim10$), very massive black holes at the centers of quasars ($z\sim6$) and massive clusters today ($z\sim1$). See for instance the recent discovery of a galaxy cluster at $z=2.5$~\cite{Wang:2016wml}. In general, massive PBH quick start structure formation, compared to the standard \LCDM paradigm, since they seed structures much more efficiently than a rare $5\sigma$-fluctuations in the Gaussian primordial spectrum~\cite{Mack:2006gz}.

\item {\bf Formation of SMBH and IMBH}. The successive PBH merging and gas accretion since recombination produces supermassive black holes (SMBH) at the centers of galaxies, and intermediate mass black holes (IMBH) at the centers of globular clusters. Their PBH origin is common to both types, and is the agent responsible for the velocity dispersion of the system (stars in the bulge and cluster respectively). Not surprisingly, there seems to be a simple power law relation between the mass of the black holes at the center of those systems and their velocity dispersion~\cite{Kruijssen:2013cna}, see Fig.~\ref{fig:IMBH}. In the MPBH scenario {\em all} DM halos have a black hole at its center, which grows by accretion of gas and the differences in masses are related to the time since first accretion. The fact that there is an underlying mechanism common to all massive black holes, from microquasars to SMBH, is seen here as a correlation which is universal.

\begin{figure}[ht]
    \begin{center}
        \includegraphics[width=12cm]{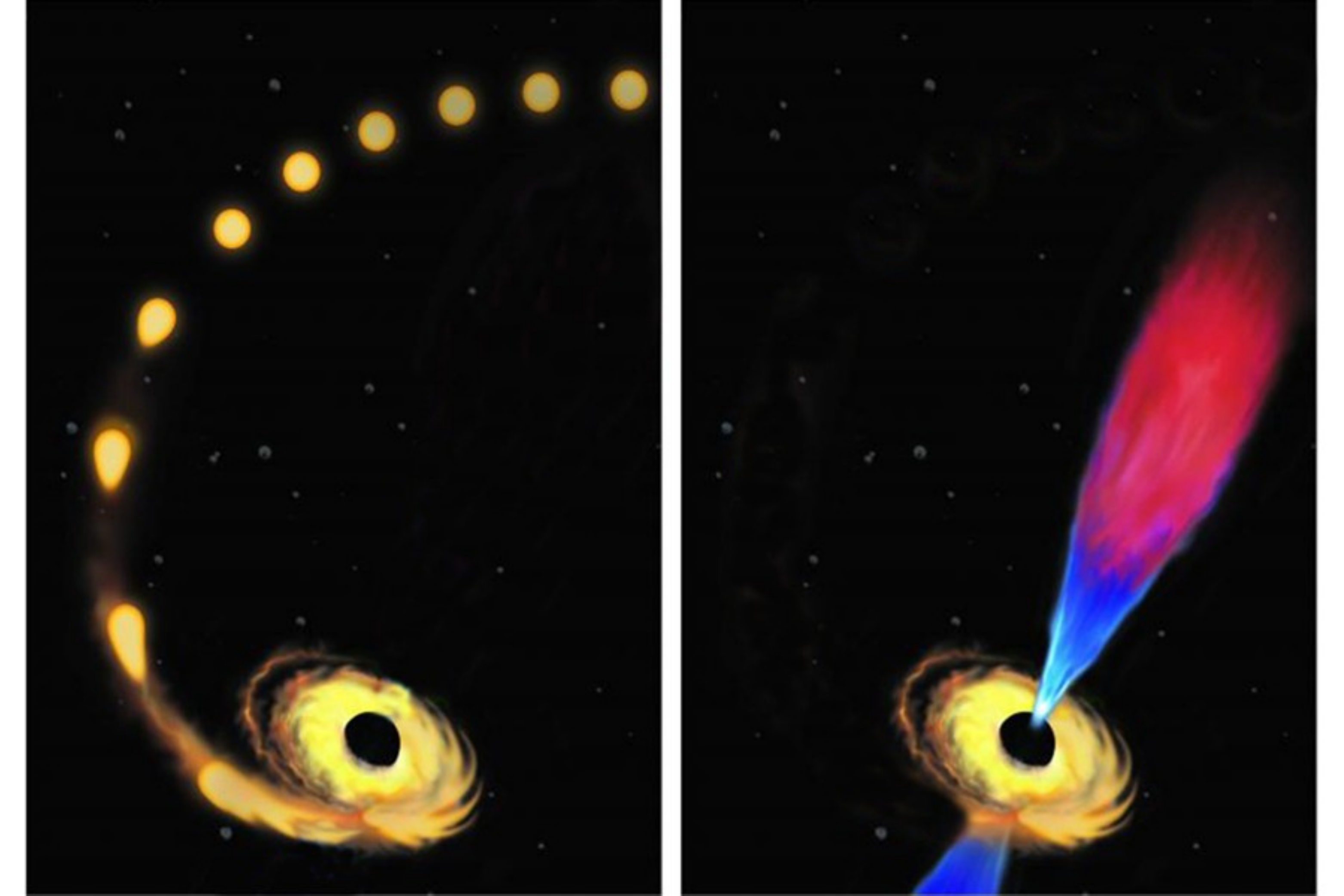}
    \end{center}
    \caption{The tidal disruption of a star captured by a PBH and subsequent flash of X-rays. This may end up on a ULX-ray transient of an accretion disk like in a microquasar. {\em Credit: modified from an original image by Amadeo Bachar}.} 
    \label{fig:BHstar}
\end{figure}

\item {\bf Superluminal SNe}. A distribution of massive PBH acting as DM in the universe should be a source of femtolensing of distant powerful sources like supernovae explosions. Some very luminous and strongly time-delayed supernovae have been observed, with emissions in the infrared that last for months before decaying. They are called superluminous SNe and there are a handful of them~\cite{Quimby:2009ps,Cooke:2012mh}. It could be that some of these are microlensed SN of known types, see e.g.~\cite{Smith:2015ucf}, whose light has been amplified by a femtolensing event on massive PBH in the line of sight.

\item {\bf Long-duration microlensing}. The duration of microlensing events towards the LMC is proportional to the mass of the lensing object~\cite{Paczynski:1985jf,Paczynski:1991zz}. If the halo of the Milky Way dark matter is constituted mainly by PBH with a range of masses peaked at around $50~\Msun$, the typical full duration is of the order of a decade. The stars in the LMC have not been monitored long enough by microlensing experiments to exclude a halo made of PBH with those masses if PBH at the peak constitute less than 10\% of the halo, see Fig.~\ref{fig:Constraints} and Refs.~\cite{LMC,Tisserand:2006zx}. Moreover, the light curves due to microlensing of clustered PBH is expected to have multiple caustics due to smaller PBHs orbiting the main deflector, and this may be difficult to distinguish from planets around stars in LMC, so they may have passed unnoticed. Also, the fact that these PBH are clustered decreases the optical depth towards a relatively compact structure like the LMC, by increasing the relative separation between clusters of PBH.

\begin{figure}[ht]
    \begin{center}
        \includegraphics[width=12cm]{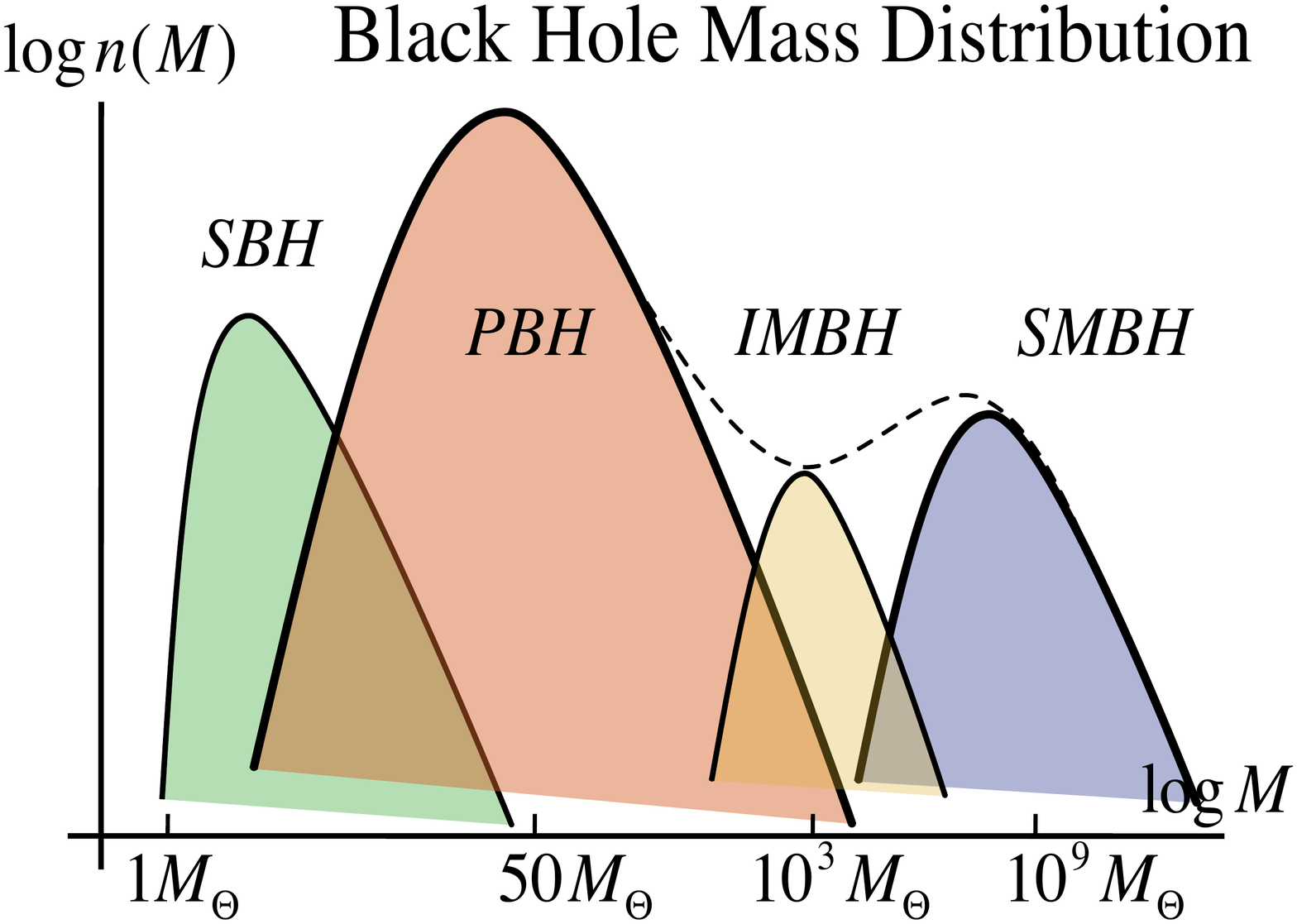}
    \end{center}
    \caption{The mass distribution of black holes in our universe can be classified in three large groups, Stellar Black Holes, that arise from the gravitational collapse of stellar systems, the Primordial Black Holes that were formed in the early Universe, the Supermassive Black Holes at the centers of Quasars and the Intermediate Black Holes at the centers of Globular Clusters. The scenario described here considers the possibility that the last three kinds are actually closely related. There could be another population of BHs at the center of microquasars with tens of solar masses and accretion disks arising from stellar capture. Their distribution is still rather uncertain, see however~\cite{Tetarenko:2016uln}. } 
    \label{fig:MassDistribution}
\end{figure}

\begin{figure}[ht]
    \begin{center}
        \includegraphics[width=12cm]{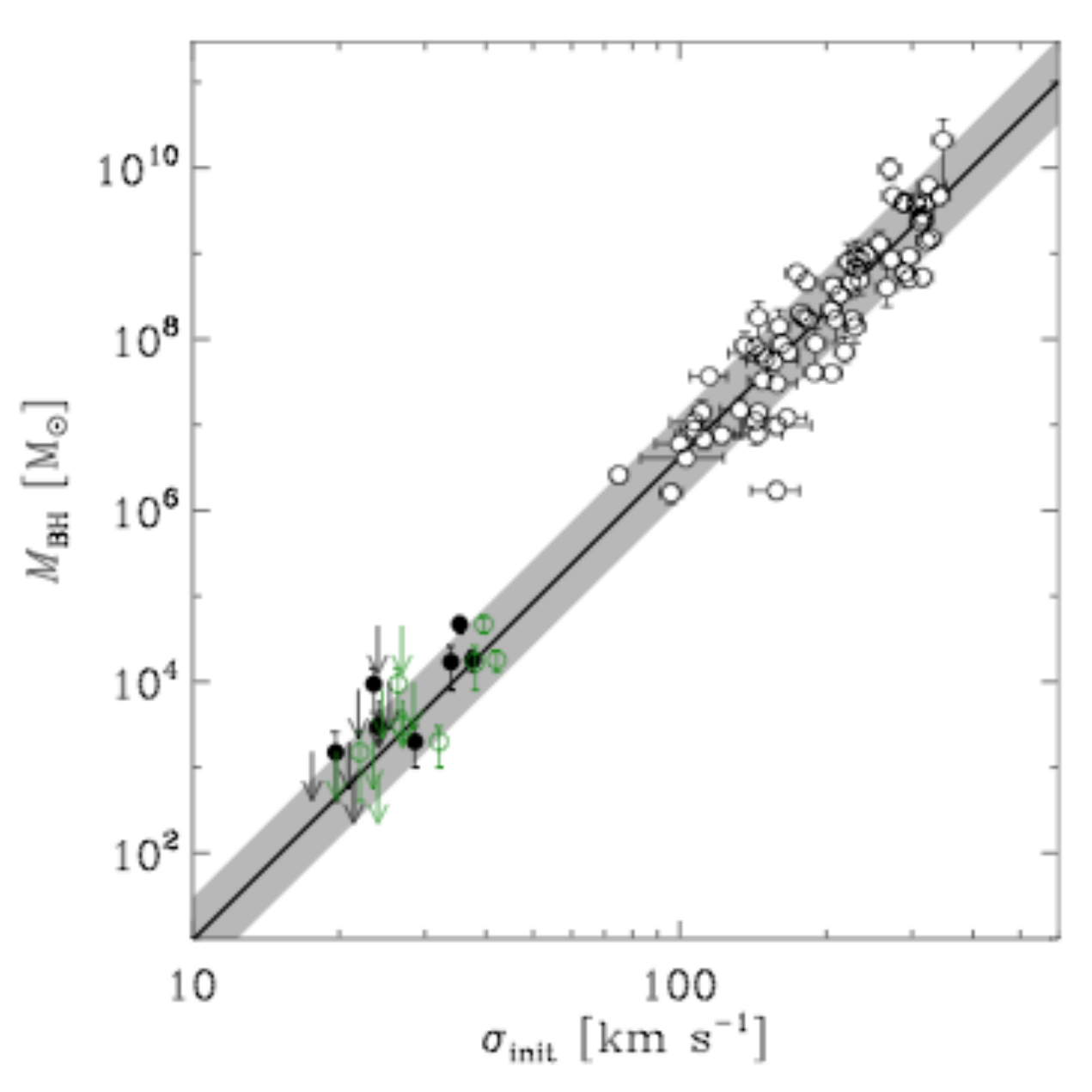}
    \end{center}
    \caption{The power law relation between the mass of the central black hole and the velocity dispersion of the system extends all the way from Supermassive Black Holes at the centers of Quasars to the Intermediate Black Holes at the centers of Globular Clusters. It is expected that further searches will show that these two classes are not distinct, but actually correspond to a common origin. In fact, a third class of BHs at the centers of microquasars in our galaxy may also have a similar relation between the central BH mass and the dispersion velocity of the gas in the accretion disk around it. From Ref.~\cite{Kruijssen:2013cna}.} 
    \label{fig:IMBH}
\end{figure}

\item {\bf Residual microlensing in time-matched strong-lensed multiple images of quasars}. Quasars that suffer strong lensing by an intervening galaxy show two or more images with identical short-time variations delayed by a constant time-shift between the images~\cite{Refsdal:1964nw}, typically of order 100 days~\cite{Dahle:2015wla}. The residual long-range modulation in amplitude has a peak and a duration above a decade, which matches a possible microlensing by a cluster of massive PBH, see also~\cite{Hawkins:2011qz}.

\item {\bf Tidally disrupted stars and ULX-ray transients}. It has been observed in both the Milky Way and nearby galaxies like Andromeda the disruption of a star by a black hole. Some of the ultraluminous X-ray transients are believed to arise from these kind of tidal-disruption events, although there seems to be more events than expected if those black holes come from stellar origin, see Fig.~\ref{fig:BHstar} and Ref.~\cite{Servillat:2013taa}.

\item {\bf Large Scale Structure N-body simulations}. The scenario of MPBH cannot be distinguished at cosmological superstructures scales from that of Particle DM, since the evolution of dark matter on large scales behaves in a way characterized by the primordial spectrum set during inflation and in accordance with the measured anisotropies in the CMB. The growth of structure occurs via the gravitational collapse of a zero-pressure fluid which, at the resolution scales of the simulations (particle mass $\sim 10^6 - 10^9~\Msun$), cannot distinguish PBH of 100 $\Msun$ from particle DM~\cite{Springel:2006vs}. 

\item {\bf Substructure and too-big-to-fail problems}. At small galactic scales, numerical simulations of structure formation based on CDM seem to give many more smaller structures than observed. We realized in~\cite{Clesse:2016ajp} that a core of MPBH at the center of shallow potential wells like those of dwarf spheroidals could be enough to expel solar-mass stars from these structures, by inducing higher than escape velocities, through the gravitational slingshot effect, see below. This could be the reason for the high mass-to-light ratios of DSph, and why we don't observe the stellar light from those substructures. These substructures have recently been observed with dedicated high exposures in sensitive cameras like DECam~\cite{DES-DSph}. Similarly with large objects that should have been seen and only recently have been detected via gravitational lensing. Thus MPBH could be responsible for the solution to the substructure and too-big-to-fail problems of Cold Dark Matter.

\item {\bf Missing baryon problem}. Perhaps PBH have eaten up most of the baryons since recombination. The last time they were measured in the CMB were in agreement with BBN light element abundance. There seems to be less in the present universe in hot plasmas around clusters and in galaxies~\cite{Maller:2004au}. They may simply not shine or they could be locked inside the PBH that account for the CDM in the Universe.

\item {\bf Cluster collisions and cross-sections}. The bullet cluster is used to constrain models of PDM by setting stringent limits on the interaction cross-sections of DM particles, $\sigma/M < 1~{\rm cm}^2/g$~\cite{Robertson:2016xjh}, which can be evaded by PBH with Schwarzschild-Born cross-sections and $M \sim 50~\Msun$, by many orders of magnitude. That is, PBHs are essentially the best candidates for collisionless dark matter.

\item {\bf Wide binaries in the Milky Way}. These are very sensitive probes of massive objects passing by, like PBHs, since they can disrupt the orbits of far-apart components of the binary and break it up. Stringent bounds have been recently updated and allow for less than a few percent halo fraction in the form of $50~\Msun$ PBH, in agreement with Fig.~\ref{fig:Constraints} and Ref.~\cite{Quinn:2009zg}.

\item {\bf Compact stellar clusters in dwarf galaxies}. Dynamical friction tend to move both massive black holes and stellar clusters to the center of dwarf galaxies, where they interact and heat up, puffing up the stellar clusters~\cite{Brandt:2016aco}. Long lived star clusters at the centers of dwarf galaxies therefore can be used to detect massive PBH, although the constraints weaken if there is an intermediate mass BH at the center of the dwarf galaxy, which provides stability and increases the velocity dispersion~\cite{Li:2016utv}.

\item {\bf Lensed Fast Radio Bursts}. Strong lensing of extragalactic fast radio bursts (FRB) induces a repetition of the burst signal with time delays of a few milliseconds. A survey of tens of thousands of sources with future radio facilities like CHIME could be used to constrain or detect MPBH in the $20 - 100~\Msun$~\cite{Munoz:2016tmg}.

\item {\bf X-ray binaries and microquasars}. We don't know the origin of X-ray binaries, where a BH accretes gas from a companion star, forming an accretion disk and emitting in X-rays. They come in two classes, low-mass and high-mass XRB. They could arise from stellar binaries that have evolved together and the companion survived the supernova explosion that lead to the BH, although typically those BHs acquire huge kicks and move away from the disk of the galaxy, most of the times loosing their stellar companion. In fact, very massive BH at the centers of accretion disks are rare outcomes from supernova explosions, since these are usually low mass BHs. It is possible to imagine a different origin of High-Mass XRB, as a companion star being captured by a passing-by isolated MPBH and initiating an accretion disk from gas outflows as the star orbits the black hole. This could be a scenario close to that of ULX-ray transients, where the impact parameter is not so small and the star survives the BH encounter loosing momentum and forming a bound system. A recent detection by Chandra of a new population of low-mass XRB~\cite{Tetarenko:2016uln} could hint in that direction. Moreover, microquasars and High-Mass XRB, that do not move at high speeds inside the galaxy, could have their origin in PBH that have captured a star and formed an accretion disk. In that case, there should be a similar relation between the central BH mass and the dispersion velocity of the gas around it, like in IMBH and SMBH, see Fig.~\ref{fig:IMBH}.

\item {\bf Fermi 3FGL and Chandra point sources}. The fourth year Fermi-LAT point source catalog shows 30\% of unknown gamma ray sources~\cite{Fermi3FGL}. Most of them are probably blazars and quasars, but many could be nearby sources within our galaxy. Some could be massive primordial black holes in dense environments that reinject energy into the ISM. The Chandra satellite also finds large numbers of compact X-ray sources in our galaxy and other nearby galaxies like Andromeda. A fraction of those could very well be MPBH. It is necessary to design methods to distinguish between the signatures of these PBH from ordinary BH from stellar evolution. There are also the EGRET third catalog of unidentified sources, many of which are in the Galaxy. Identifying the spatial distribution of those sources may give a clue as to their origin.

\item {\bf GAIA anomalous astrometry}. If the halo of dark matter in our galaxy is not made up of a diffuse gas of particles but rather of compact objects like MPBH, one would expect close encounters of stars with those hard cores, such that individual stellar motions will be affected, inducing anomalous deviations with respect to their surroundings. A careful monitoring of positions and velocities of billions of stars like that planned by GAIA~\cite{GAIA} will be able to discover those anomalous motions and predict the location and velocities of those MPBH in our galaxy that constitute the DM.

\item {\bf Dynamical friction towards the center}. Massive objects tend to concentrate via dynamical friction towards the center of potential wells. MPBH within a range of masses will thus tend to accumulate at the centers of all intermediate structures, from globular clusters, through dwarf spheroidals, to our own galaxy. In particular one expects MPBH to merge and form more massive IMBH, which will tend to orbit today around the SMBH at center of all galaxies. There is a chance that we may discover some IMBH around the SMBH at the center of our galaxy, the Milky Way, see e.g.~\cite{Oka:2016}.

\item {\bf Emission of GW in PBH binaries (LIGO)}. In their dynamical evolution as a pressureless fluid, MPBH will naturally merge emitting gravitational waves like those recently detected by LIGO. Now we know that there has been at least three GW merging events detected by LIGO in three months. Their masses range from $60~\Msun$ (after merging) to $8~\Msun$ and therefore cannot be described by a monochromatic spectrum of PBH masses, rather by a wide mass distribution. It is true that three events are not enough to characterize the whole mass distribution, but more events have already been reported by LIGO in Run O2, so we expect within a decade to have enough events to make statistical claims~\cite{Clesse:2016vqa}.

\item {\bf Mass and spin distribution of PBH}. For the moment, we are probably only detecting with LIGO the tip of the MPBH distribution, where the majority of the PBH live, but soon we will have of the order of one BBH coalescence per day, with the expected advanced LIGO nominal sensitivity, which will allow us to measure their mass distribution with a few percent accuracy. Unfortunately, due to the seismic noise barrier of terrestrial interferometers, we may not be able to map well the mass distribution of PBH between 100 and $10^4~\Msun$, since Advanced LIGO is sensitive only to masses $M_{\rm PBH} < 150~\Msun$, due to strain sensitivities above 30~Hz and a mass-dependent ISCO frequency, $f_{\rm ISCO} = 44~{\rm Hz}~(100\Msun/M)$, see Fig.~\ref{fig:LIGOevents}. Larger masses could only be probed by future interferometers like Einstein Telescope. Moreover, with better BBH merging templates and GW detections, we will be able to measure not just their mass distribution, but also the spin distribution. Together this will help us distinguish the origin of these BBH mergers. Primordial BHs naturally form by gravitational collapse of a spherical overdensity the size of the horizon when a primordial fluctuation reentered during radiation epoch, and therefore it is expected that PBH were born without spin, in a cluster of many more PBH which merge successively in pairs. It is expected that the spin of the resulting black hole from each merging will have a non-negligible spin, although by the time several isotropic mergings have occurred, those spins could be averaged out. Future N-body simulations will be performed to study this issue.

\item {\bf Stochastic background of GW (LISA)}. The merging of PBH soon after recombination to form larger mass black holes is a powerful source of gravitational waves, with efficiency of order few permil, which redshifts faster than matter and thus can reach $\Omega_{\rm GW}h^2 \sim 10^{-6}$ today, from unresolved mergers, going back to at least $z\sim200$, and frequencies in a broad range, from the nanoHertz, possibly detectable by pulsar timing arrays, through miliHertz, easily detectable by the proposed LISA satellite~\cite{Bartolo:2016ami,Clesse:2016ajp}, down to kiloHertz seen by terrestrial interferometers like LIGO, VIRGO, KAGRA, etc. There is also the stochastic background generated at PBH formation. The violent gravitational collapse to form PBH at horizon reentry produces also a small fraction of energy, of order $10^{-6}$ of the total, in the form of gravitational waves, or $\Omega_{\rm GW}h^2 \simeq 10^{-11}$ today. 

\end{itemize}

\begin{figure}[ht]
    \begin{center}
        \includegraphics[width=12cm]{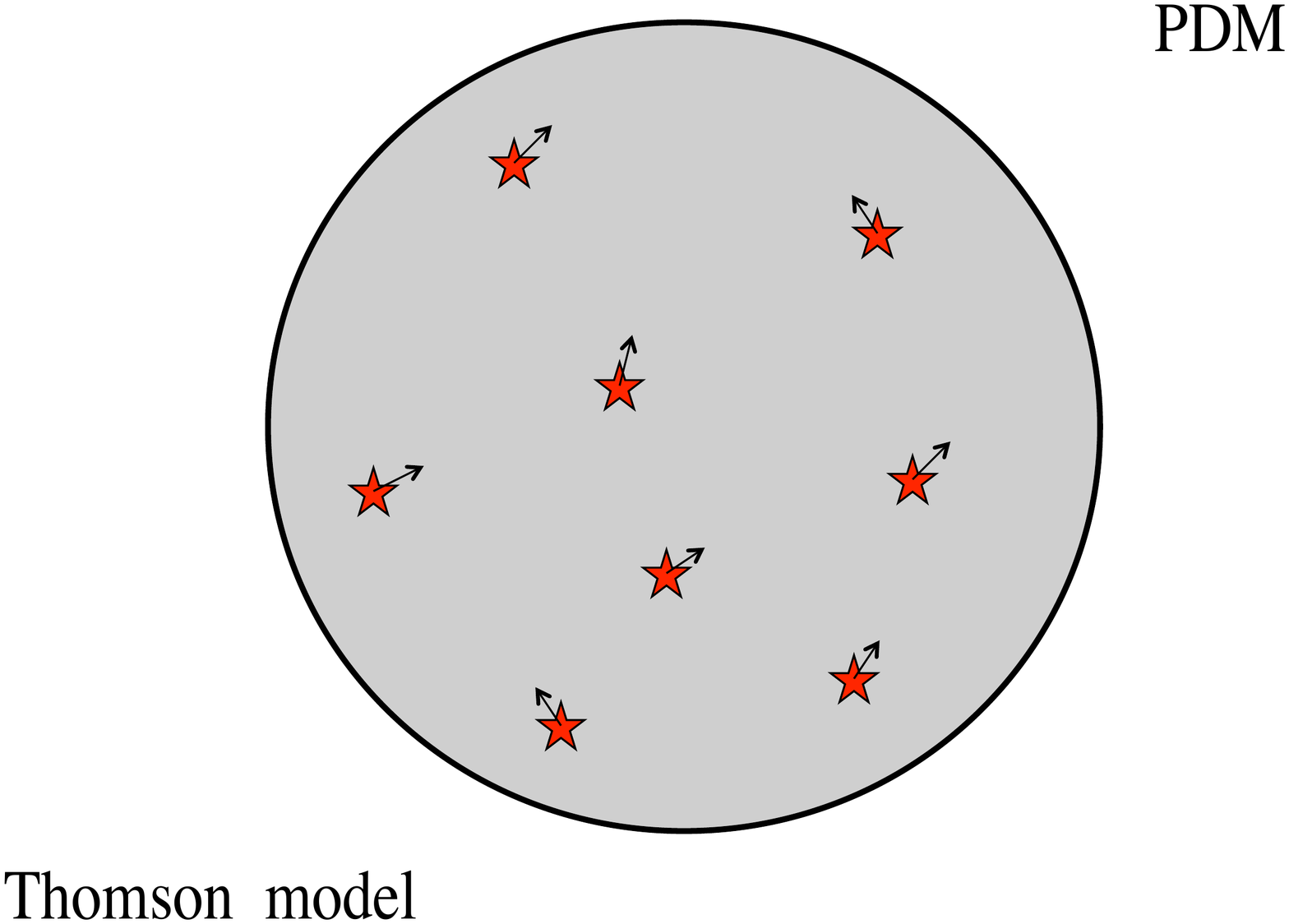} \\
        \includegraphics[width=12cm]{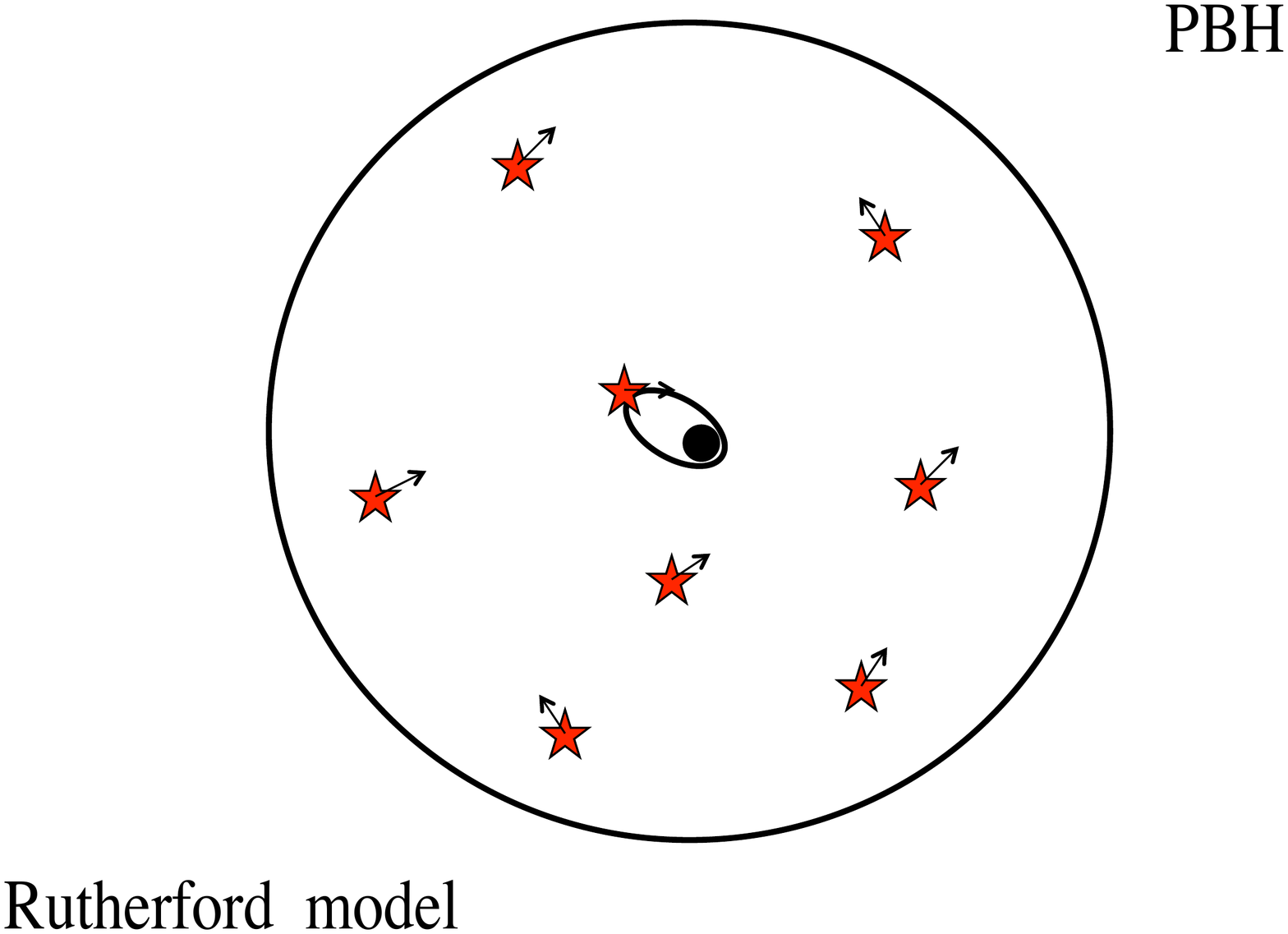}
    \end{center}
    \caption{The standard paradigm of particle dark matter assumes a uniform distribution of DM particles around stars. This new PBH scenario substitutes a large volume for a compact central massive object. There is a natural analog between the PDM scenario and the Thomson model of the atom and the PBH scenario and the Rutherford model. While stars move in the PDM halo without any effort, in the new scenario a few stars may have close encounters with the PBH at the center and abruptly change their trajectories.} 
    \label{fig:ModelPDM}
\end{figure}

\section{Anomalous motion of stars}
\label{sec:Anomalous}

\

In this section we describe the effect of PBH on the nearby starts, as could be measured by the satellite GAIA, as well as the effect on stars of shallow potential wells of dwarf spheroidals, which could have acted by expelling stars from them via the gravitational slingshot effect.

\begin{figure}[ht]
     \vspace*{-5mm}
    \begin{center}
        \includegraphics[width=12cm]{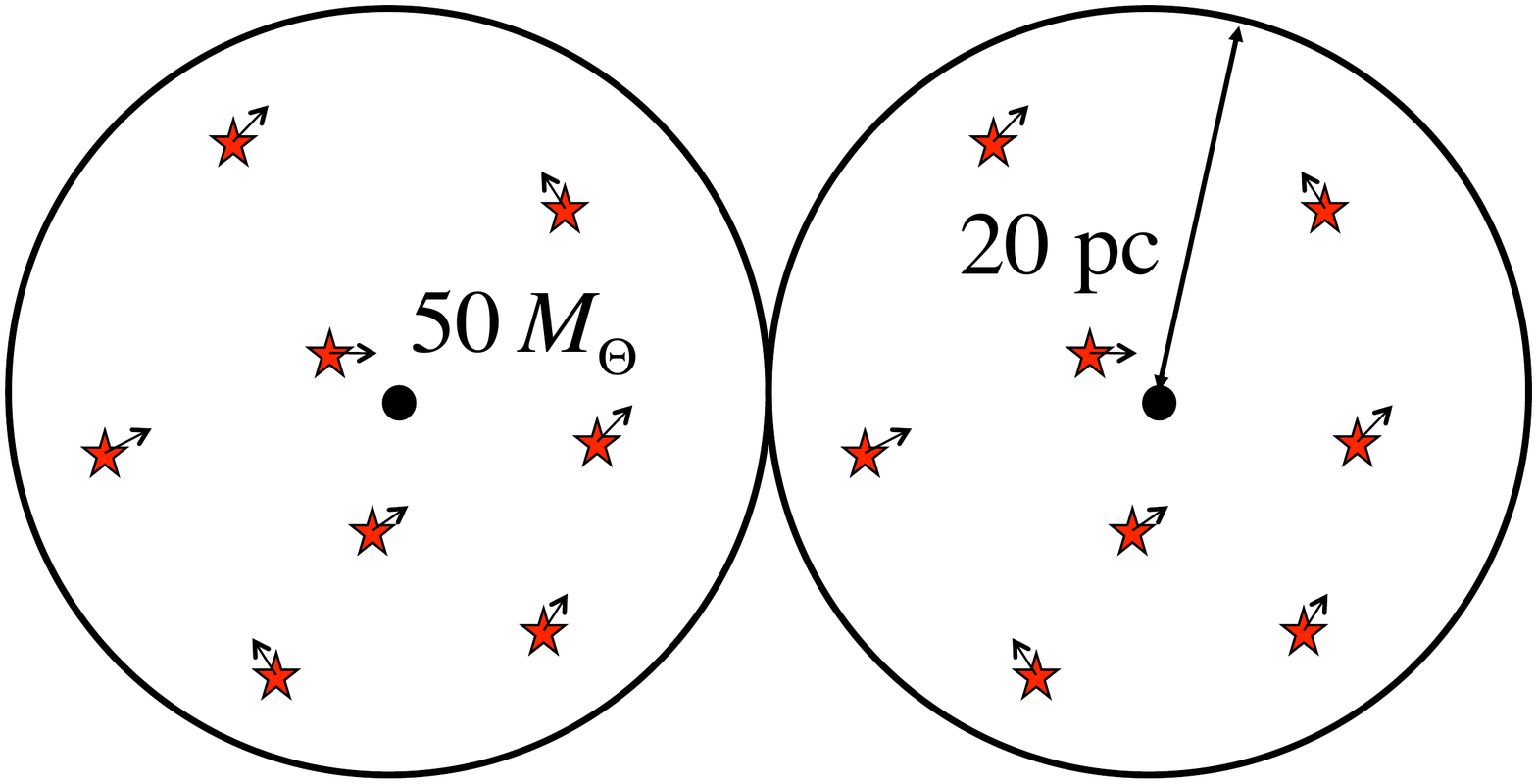}
    \end{center}
     \vspace*{-10mm}
    \caption{The average distance between PBH depends very much on their clustering. Assuming PBH form the bulk of the DM in the halo, we can compute their typical separation. For isolated PBH, of mass $M\sim50~\Msun$, the typical distance is $40$~pc. Such a volume comprises close to 100,000 stars separated approximately 1\,pc from eachother. Only a few stars in $\sim 10^5$ will have close encounters (of less than a parsec) with the compact PBH forming the halo.} 
    \label{fig:MeanDensity}
\end{figure}

\subsection{GAIA and anomalous astrometry}

\

We first compute the anomalous acceleration, velocity and displacement induced by a population of PBH in the solar neighborhood. The dark matter density near the Sun was measured recently by Ref.~\cite{Kafle:2014xfa} 
\be
\rho_\odot^{\rm DM} = 0.0088 \pm 0.0020 \,\Msun/{\rm pc}^3\,,
\ee
which gives a typical distance between black holes of 
\be
d_{\rm PBH} \equiv 2\times
\left(\frac{\rho_{\rm halo}}{M_{\rm PBH}}\right)^{-1/3} =
40\,{\rm pc} \, \left(\frac{M_{\rm PBH}}{50~\Msun}\right)^{1/3}\,.
\ee
This means that in a halo volume with a single PBH of $50~\Msun$ in the solar neighborhood, there are about $10^5$ stars, separated from each other typically by 1~pc. Most of these stars will not feel the presence of the nearby PBH. The gravitational attraction on each of these stars is given by
\be
a_i = \frac{GM_{\rm PBH}}{b^2} + \sum_j \frac{Gm_j}{r_{ij}^2}\,\hat u_j \,,
\ee
where the first term is due to the MPBH at a distance $b$ from the star,\footnote{For a fluid of dark matter particles permeating the solar neigborhood this term would be absent by Gauss law.} and the second term is due to the rest of the stars in the neighborhood. The presence of a dark compact object like a MPBH can be deduced from the {\em anomalous} acceleration induced by the MPBH,
\be
\Delta a = 1\times 10^{-15}\,m\,s^{-2}
 \left(\frac{M_{\rm PBH}}{50~\Msun}\right)
  \left(\frac{20\,{\rm pc}}{b}\right)^2\,,
\ee
which seems utterly negligible. However, for impact parameters $b < 0.1$~pc, which may happen for one in $10^5$ stars in the field, one can reach significant accelerations. The anomalous relative velocity induced in those cases is
\be
\frac{\Delta v}{v_0}(0.1\,{\rm pc}) \sim 3\times 10^{-4}\,,
\ee
for typical halo velocities of $v_0 \sim 200$ km/s. These induce anomalous displacements which, at a distance of approximately $x_0\sim 1$~kpc, could be seen as transverse displacements of order
\be
\frac{\Delta x}{x_0}(0.1\,{\rm pc}) \sim 0.56~{\rm marcsec}\,,
\ee
which are in principle detectable with GAIA~\cite{GAIA}.

\begin{figure}[ht]
    \begin{center}\vspace*{-6mm}
        \includegraphics[width=12cm]{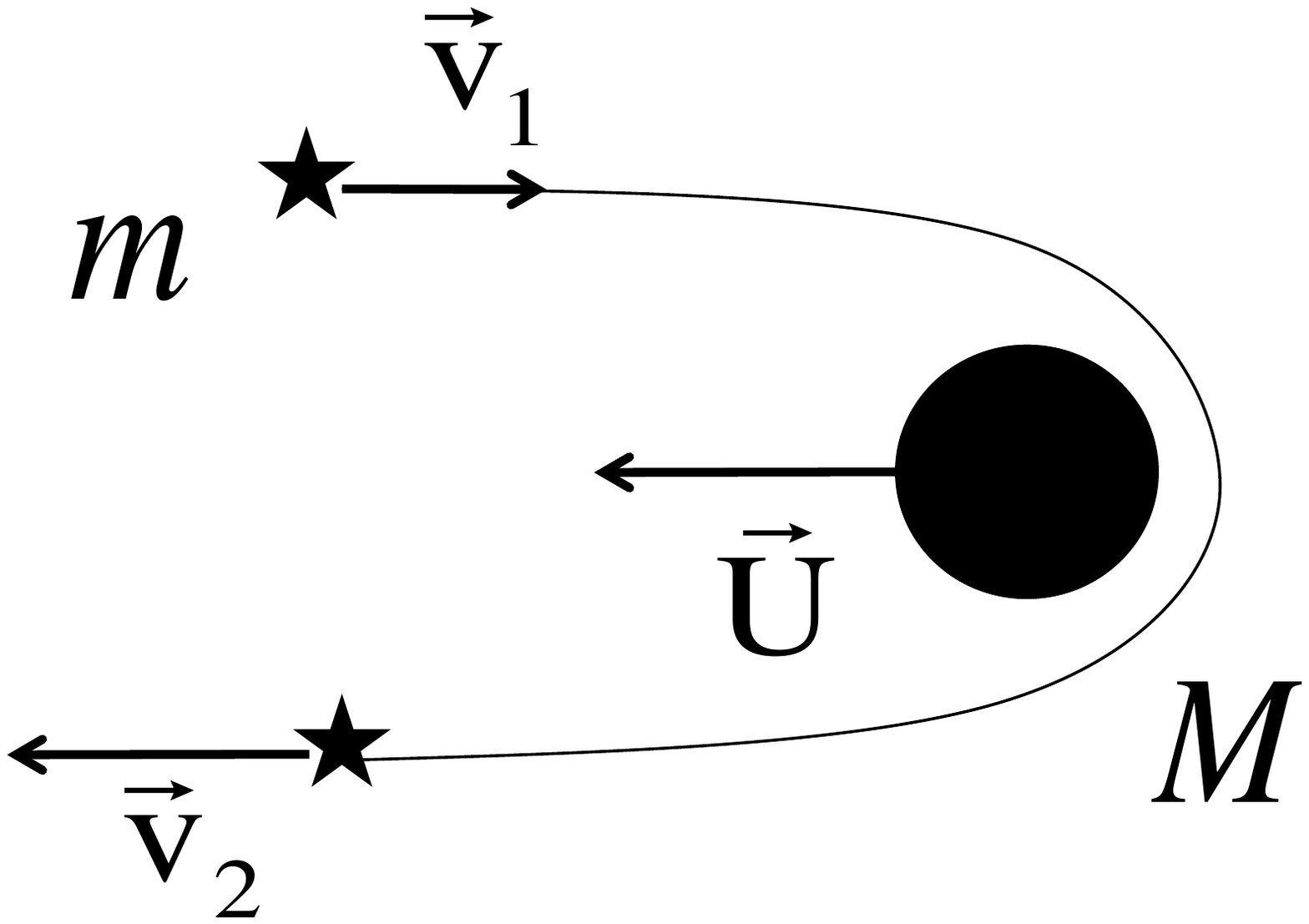}
     \vspace*{-15mm}
    \end{center}
    \caption{The gravitational slingshot effect converts a small peculiar velocity $\vec{v}_1$ of a star of mass $~m~$ into a larger than escape velocity $\vec{v}_2$, thanks to the exchange of energy and momentum with the MPBH of mass $M$.} 
    \label{fig:Slingshot}
\end{figure}

\subsection{Ejecting stars from Globular Clusters and DSph with the
gravitational slingshot effect}

\

Apart from anomalous motion of stars in our solar neighbourhood, one can also compute the effect of close encounters of stars with massive PBH in globular clusters and dwarf spheroidals with shallow potential wells, where the typical velocities of stars are tens of km/s, while those of MPBH (at the core of the potential well, due to dynamical friction) are close to a hundred km/s. Thanks to energy-momentum conservation, stars that pass close to these MPBH will be expelled from the cluster. They will suffer the well known gravitational slingshot effect: a star of mass $m$ with incoming velocity $\vec v_1$ encounters a MPBH with mass $M$ and velocity $\vec U$, and through gravitational recoil the outgoing star acquires a velocity $\vec v_2$, see Fig.~\ref{fig:Slingshot}, with magnitude given by
\be\label{recoil}
v_2 = \frac{2U + (1-q)\,v_1}{1+q}\,,
\ee
where $q=m/M$. If the MPBH has a large velocity (since it probes the core of the potential well) then it will imprint on the star a velocity twice as large. If the star and the black hole encounter each other at an angle $\theta$, the recoil velocity will be slightly different, 
\be\label{recoiltheta}
v_2(\theta) = \sqrt{v_1^2\sin^2\theta + \left(\frac{2U + (1-q)\,v_1\cos\theta}{1+q}\right)^2}\,,
\ee
but marginalizing over angles,
\bea\label{recoilmarg}
\bar v_2 &=& \frac{1}{2}\int_0^\pi d\theta\,\sin\theta\ v_2(\theta) = \nonumber \\
&=& \frac{v_1}{8}\left[ -\frac{2\beta(1-6\alpha+\alpha^2)}{1+\alpha} + \frac{(1+\alpha)(1+\alpha\beta^2)}{\sqrt\alpha}
\times \right. \nonumber \\
&&\hspace{-5mm} \left.\left(\arctan\frac{\sqrt\alpha(2+\beta-\alpha\beta)}{2\alpha\beta+\alpha-1} + \arctan\frac{\sqrt\alpha(2-\beta+\alpha\beta)}{2\alpha\beta-\alpha+1}\right)\right] \nonumber \\
&\simeq& \frac{2U}{1+q}\,,
\eea
which is almost identical to Eq.~(\ref{recoil}), for $\alpha = 1/q \gg 1$, and $\beta = U/v_1\gg1$.

With a few of these BH bypasses (slingshots), the star will acquire a velocity above the escape velocity of the globular cluster or the dwarf spheroidal, and be expelled from the shallow potential well.\footnote{A similar mechanism has been invoked in Ref.~\cite{Levin:2005hs} to explain the ejected stars from the galactic center by the presence of an IMBH~\cite{Hansen:2003yb}.} This may be the reason why small substructures like dwarf galaxy satellites, with masses below $10^6$ to $10^8~\Msun$, do not shine: they could have lost most of their stars by ejections, and thus present today large mass-to-light ratios, of order 300 to 1000, as recently measured~\cite{DES-DSph}. This may explain both the substructure and the too-big-to-fail problems of standard CDM scenarios.

\section{Microlensing within strong lensing of QSO}
\label{sec:QSO}

\

Well known quasars like QSO 0957+4561 have two or more images, whose short-time magnitude pattern matches, once we take into account a constant time delay due to the different paths that light has travelled from each source. However, the residual long-time pattern is not constant, but follows a characteristic amplification light curve of more than a decade~\cite{Hawkins:2011qz}, which suggests that one of the lines of sight has been microlensed by a black hole of a few tens of solar masses. Detailed studies of several of these strong-lensed quasars with multiple images may allow us to determine not only the rate of expansion~\cite{Refsdal:1964nw}, from the time delay between images~\cite{Dahle:2015wla}, but also the dark matter distribution through the long-time residuals, if these happen to show microlensing amplification~\cite{Hawkins:2011qz}.

\section{Gamma and X-ray emission}
\label{sec:Gamma}

\

Ever since satellites have explored the sky in X-rays and gammas, they have discovered multiple bright sources, some point-like and others more extended, and even a diffuse background. Some of the sources have been identified with  distant AGN and QSO sources, outside our galaxy. Others are neutron stars and X-ray binaries. There is, however, a growing list of unidentified sources in all point source catalogs, from EGRET to Chandra/XMM and Fermi LAT. Some of these sources are extragalactic, but many are unknown galactic sources, with hard spectra. There is the natural assumption that at the core of these sources lie a black hole, some more massive than others, but in any case there seems to be too many BH to be accounted for by the standard stellar evolution and subsequent supernova explosions. Extrapolated to the history of the galaxy, it would imply that there should have been a larger than observed cosmic ray flux and thus a larger UV and IR background.

\begin{figure}[ht]
    \begin{center}
        \includegraphics[width=12cm]{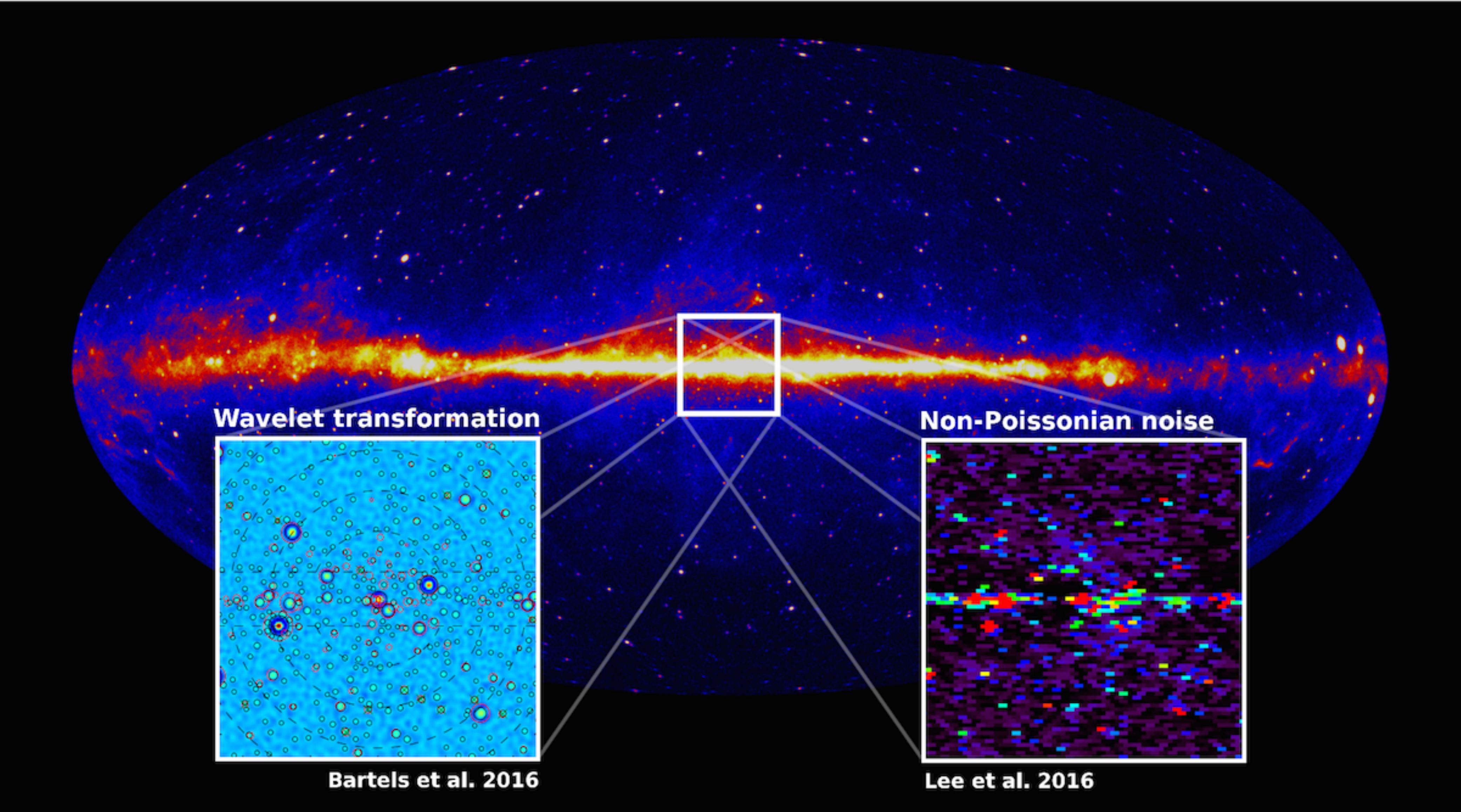}
    \end{center}
    \caption{The Fermi 3FGL point source catalog presents a large number of unidentified sources, as well as a diffuse background that could be consistent with bright point sources due to gamma emission around PBH. Figure from CERN Courier~\cite{Courier}.} 
    \label{fig:Fermi}
\end{figure}

Moreover, the recent Fermi LAT measurements of the diffuse gamma-ray background towards the galactic center (GC) has an uncertain origin. Some researchers speculated that it could have arised from PDM annihilations, but this has been discarded as a reasonable explanation by two independent groups~\cite{Bartels:2015aea,Lee:2015fea}, which have found that the diffuse GC background is better described by a population of unresolved point sources, adding extra support to compact massive objects like black holes as the more plausible origin of the GC background.

It is therefore worth studying these individual point sources, e.g. in the 3FGL catalog of Fermi-LAT, to probe their nature and determine whether a PBH could be their source of energy, rather than a pulsar or a massive star. Some of these sources are correlated with molecular clouds close to the galactic center and these shine in X-rays and many other wavelengths. Moreover, since the spatial distribution of PBH should follow that of the bulk of the dark matter in the halo, it is expected that some spatial correlation between the X-ray and gamma-ray sources of EGRET, Chandra and Fermi and a generic NFW or Einasto profile should exist. Furthermore, since the power spectrum of  PBH dark matter has a broad peak, responsible for the clustering of the initial primordial black holes, one should expect to see higher spatial correlations at short distances of order a parsec or smaller.

\section{Gravitational wave emission}
\label{sec:GWE}

\

The MPBH scenario has three different sources of gravitational waves at very different frequencies. The broad and high peak in the curvature fluctuation power spectrum is generated during inflation from the slow-roll motion of the inflaton field and possibly other fields coupled to it. Scalar fluctuations have a large enough amplitude that backreact on both scalar and tensor modes, creating a sizeable gravitational wave background with a peak at frequencies associated with the size of the horizon at reentry that will have redshifted today. Then there is the GW stochastic background from the formation of PBH themselves. The gravitational collapse that gives rise to the PBH at horizon reentry during the radiation era produces also a small fraction of energy, of order $10^{-6}$ of the total, in the form of gravitational waves, which have redshifted (from the time of formation at $z=z_f$) to $\Omega_{\rm GW}h^2 \simeq 10^{-11}$ today. Such a background is similar to that produced at violent first order phase transitions, and will have a peak at a typical frequency today 
$$f^{\rm peak} \simeq 1.4\times10^{-8}\,Hz \left(\frac{30\,\Msun}{M_{\rm PBH}^{\rm peak}}\right)^{1/2}\,,$$ 
and an amplitude at the peak given by
$$h_c(f^{\rm peak}) \simeq 2.85\times10^{-16} \left(\frac{M_{\rm PBH}^{\rm peak}}{30\,\Msun}\right)^{1/2}\,,$$
for PBHs of mass peaked at
$M_{\rm PBH}^{\rm peak} \simeq 30\,\Msun \left(\frac{7\times10^{11}}{1+z_f}\right)^2$\,,
which falls in the range of sensitivity of future Pulsar Timing Arrays.

\subsection{Stochastic background of GW}

\

\begin{figure}[ht]
    \begin{center}
    \hspace*{-7mm}
        \includegraphics[width=13cm]{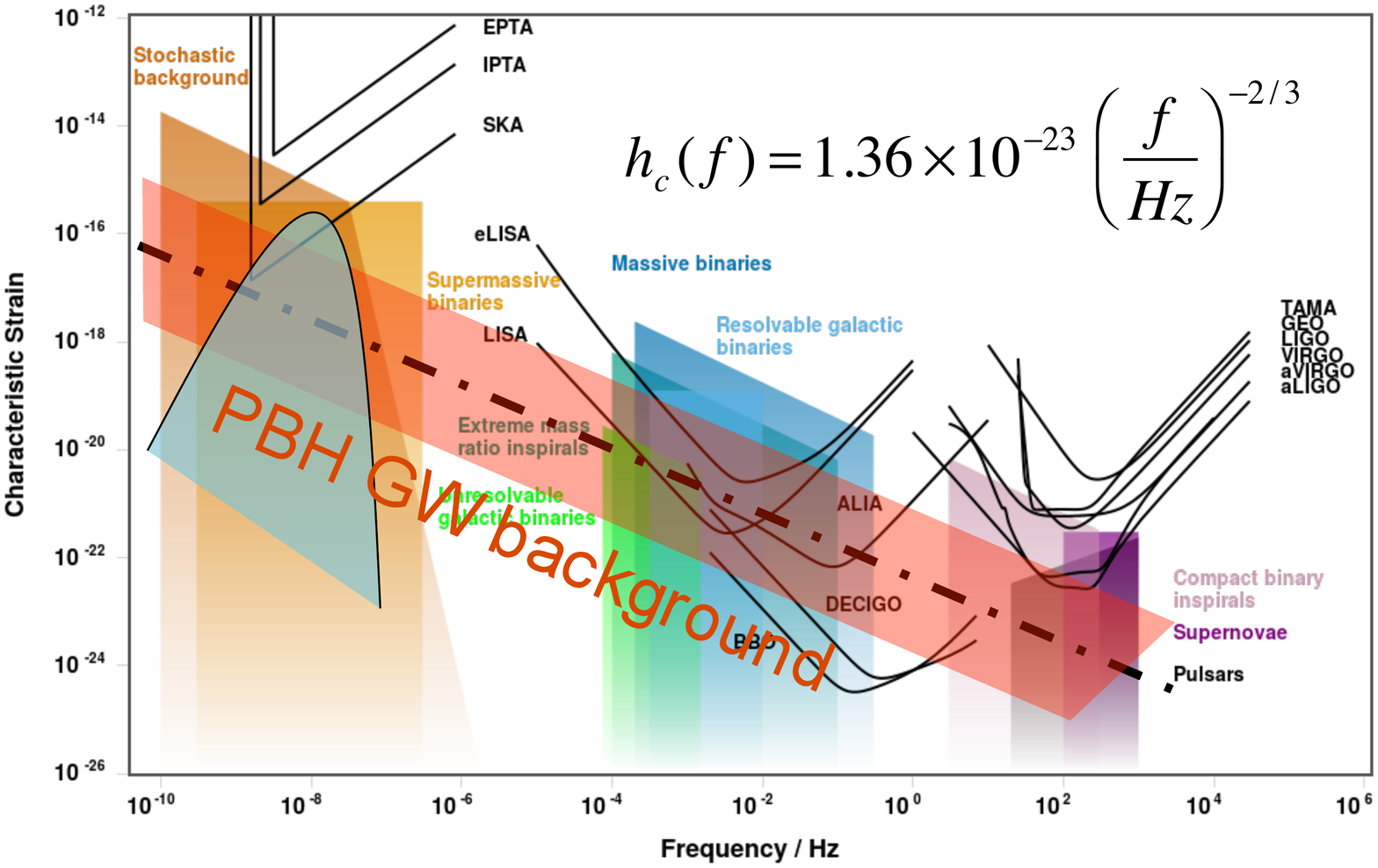}
     \vspace*{-10mm}
    \end{center}
    \caption{The stochastic background of gravitational waves from the merging of primordial black holes since recombination spans many orders of magnitude, from below the nanoHertz to above the kiloHertz. The range is covered by ground interferometers, satellites and pulsar timing arrays. Also shown in grey is the stochastic background of gravitational waves created from the violent gravitational collapse that formed the PBH upon horizon entry of large fluctuations during the radiation era.  Adapted from Ref.~\cite{GWplotter}.} 
    \label{fig:Stochastic}
\end{figure}

Finally there is the GW stochastic background arising from the inspiralling and merging of PBH during the matter era and detected as unresolved sources by LIGO, LISA and PTA. For a constant rate of merging since recombination at $R\sim 55$ events/yr/Gpc${}^3$~\cite{TheLIGOScientific:2016wyq,TheLIGOScientific:2016pea}, in a \LCDM universe, one finds~\cite{Clesse:2016ajp}
\be
h_c(f) \simeq 1\times10^{-24} \left(\frac{{\cal M}_c}{\Msun}\right)^{5/6}\!\left(\frac{f}{Hz}\right)^{-2/3}\,,
\ee
where ${\cal M}_c^{5/3} = M_1M_2(M_1+M_2)^{-1/3}$ is the chirp mass, and
\be
h^2\Omega_{\rm GW}(f) \simeq 8\times10^{-13} \left(\frac{{\cal M}_c}{\Msun}\right)^{5/3}\!\left(\frac{f}{Hz}\right)^{2/3}\,,
\ee
which is just below the Advanced LIGO sensitivity but could eventually be detected by LISA, at lower frequencies, see Fig.~\ref{fig:Stochastic}. 

\begin{figure}[ht]
    \begin{center}\hspace*{-6mm}
        \includegraphics[width=13cm]{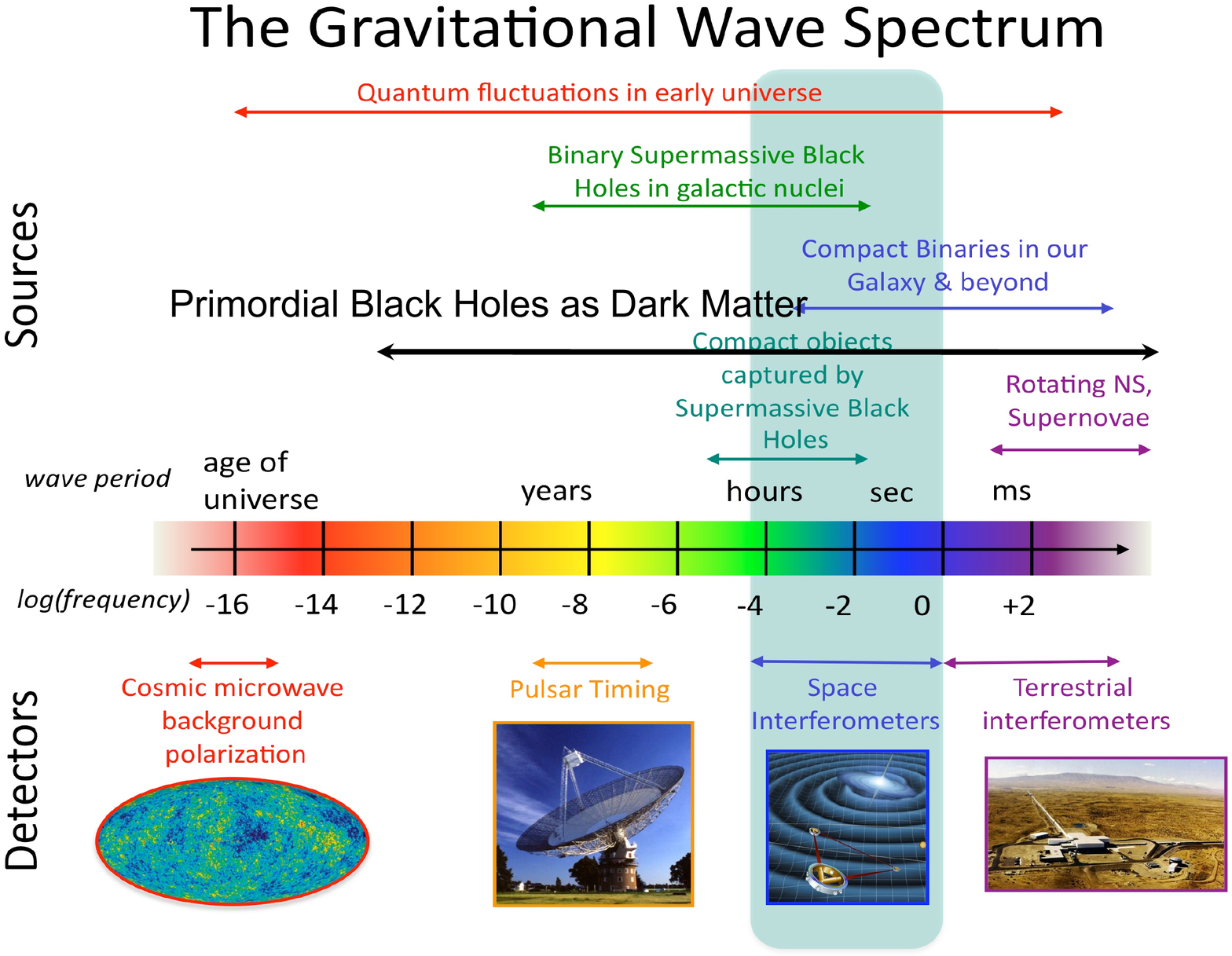}
     \vspace*{-5mm}
    \end{center}
    \caption{The spectrum of gravitational waves from the merging of primordial black holes spans many orders of magnitude, from below the nanoHertz to above the kiloHertz. The range is covered by ground interferometers, satellites and pulsar timing arrays.  {\em Credit: modified from an original image by NASA Goddard Space Flight Center}.} 
    \label{fig:GWspectrum}
\end{figure}

The spectrum of gravitational waves from MPBH thus covers a huge range of frequencies and amplitudes, see Fig.~\ref{fig:GWspectrum}. Some may be accessible through terrestrial GW interferometers, in the kHZ domain, others through GW antennas in space, in the mHz, and yet another by Pulsar Timing Arrays, in the nHz range.

\section{Discussion and Conclusions}
\label{sec:Discussion}

\

I have presented here a new paradigm, based on massive PBH, rather than fundamental particles, as the main component of Dark Matter. MPBH could arise from peaks in the spectrum of matter fluctuations produced during inflation. With the future terrestrial and space GW interferometers we may be able to characterize their distribution and properties, providing clues about their origin and the early stages of the evolution of the Universe. Those MPBH have typically a broad mass distribution, from below a solar mass to several tens of solar masses. Large masses are difficult to detect with the present GW interferometer LIGO, due to the seismic wall at low frequencies. However, if in the near future a black hole is detected by LIGO in a Compact Binary Coalescense with a mass smaller than solar, it could never have been produced by stellar collapse, due to the fundamental Chandrasekhar limit, and therefore is a clear signal of primordial origin.

There is already a lot of evidence in favor of PBH as DM. The strong correlation between fluctuations in the infrarred and the diffuse X-ray backgrounds suggests that a large population of PBH could have initiated early structure formation and reionization, and at the same time be responsible for the present X-ray background. This speed-up in structure formation may explain the existence of fully formed galaxies and clusters at high redshift, much before expected from the standard $\Lambda$CDM paradigm. Moreover, PBH could also be responsible for the observed ultra luminous x-ray sources, the SMBH and IMBH at centers of galaxies and globular clusters, and alleviate the substructure and too-big-to-fail problems of the standard CDM paradigm.

Moreover, in this work I describe multiple signatures within a large variety of phenomena in Astrophysics and Cosmology where the presence of MPBH have an impact. Some are well known, others are new avenues of exploration. Here is a comprehensive list:
CMB distortions and anisotropies;
First stars in the Universe;
Reionization and 21cm Intensity mapping;
Early galaxy formation;
SMBH and IMBH at the centers of galaxies and globular clusters;
Superluminal SNe;
Long-duration microlensing;
Residual microlensing in time-matched strong-lensed multiple images of quasars;
Tidally disrupted stars and Ultra Luminous X-ray transients;
Substructure and too-big-to-fail problems in LSS N-body simulations;
Missing baryon problem;
Cluster collisions and cross-sections;
Wide binaries in the Milky Way;
Compact star clusters in dwarf galaxies;
Lensed Fast Radio Bursts;
X-ray binaries and microquasars;
Fermi 3FGL and Chandra point source catalogs;
X-ray and Gamma Diffuse backgrounds;
GAIA anomalous astrometry;
Dynamical friction and formation rate of SMBH;
Emission of GW in MPBH binaries;
Mass and spin distribution of PBH;
Stochastic background of GW. 
All of them are not only tests of the new paradigm, but also opportunities to constrain the Early Universe scenario behind the MPBH origin.

The fact that quantum fluctuations of the inflaton field can backreact on space-time and form classical inhomogeneities, giving rise to CMB anisotropies and big structures like galaxies and clusters, is a fascinating property of Quantum Field Theory in curved space, and one of the great successes of inflation. What is even more surprising is that a local feature in the inflaton dynamics can give rise to large amplitude fluctuations in the spatial curvature, which collapse to form black holes upon reentry during the radiation era, and that these PBH may constitute today the bulk of the matter in the universe. In this new scenario, Dark Matter is no longer a particle produced after inflation, whose interactions must be deduced from high energy particle physics experiments, but an object formed by the gravitational collapse of relativistic particles in the early universe when subject to high curvature gradients, themselves produced by quantum fluctuations of a field and stretched to cosmological scales by inflation.

In conclusion, Massive Primordial Black Holes are the perfect candidates for collisionless CDM, in excellent agreement with present observations. This new scenario opens up a very rich phenomenology that can be tested in the near future from CMB, LSS, X-rays and GW observations, in a multi-probe, multi-epoch and multi-wavelength approach to study the nature of Dark Matter. Furthermore, if MPBH are indeed responsible for DM, and its properties are measured, it will open a new window into the Early Universe, where the dynamics occurring in the last stages of inflation may be probed by astrophysical and cosmological observations.

\

\section*{Acknowledgments}

\

I thank Sebastien Clesse, Andrei Linde, Marco Peloso, Ester Ruiz Morales, Caner Unal and David Wands, for a very fruitful collaboration that has led to the massive PBH scenario described here. These proceedings are a transcription of the various talks I have given during 2015 and 2016 in several workshops and conferences, with further insights and new research avenues. I also thank Lam Hui, Alan Heavens, Miguel Zumalac\'arregui, Ruth Durrer, Alvise Raccanelli, Lenny Susskind, Peter Michelson, Peter Graham, Javier Rico, Manel Martinez, Valenti Bosch-Ramon, Emma de O\~na, German Vargas, Savvas Nesseris, Jos\'e Pedro Mimoso, Julian Mu\~noz, Jim Annis, Tamara Davis, Alex Drlica-Wagner, Alexandre Refregier, Mikhail Shaposhnikov, Marti Raidal, Kfir Blum and many others for very insightful discussions on the physics of PBHs.
This work is supported by the Research Project of the Spanish MINECO FPA2015-68048-C3-3-P and the Centro de Excelencia Severo Ochoa Program SEV-2012-0249.

\section*{References}

\

\end{document}